\documentclass[%
 notitlepage,
superscriptaddress,
 amsmath,amssymb,
 aps,
 reprint,
pra,
nofootinbib
]{revtex4-2}
\usepackage[dvipsnames]{xcolor}

\usepackage{graphicx}% Include figure files
\usepackage{dcolumn}% Align table columns on decimal point
\usepackage{bm}
\usepackage{hyperref}
\hypersetup{colorlinks = true, linkcolor = [rgb]{0.19411,0.51882,0.667058}, urlcolor = [rgb]{0.125490,0.29542,0.1647058}, citecolor = [rgb]{0.75882,0.37411,0.14117}}

\usepackage{setspace}

\usepackage{amssymb}

\usepackage{braket}

\graphicspath{{../images/}}
\usepackage[toc,page]{appendix}

\newcommand{\nn}{\nonumber \\}

\usepackage{braket}

\def\>{\rangle}
\def\<{\langle}

\usepackage[bottom]{footmisc}

\usepackage{setspace}
\usepackage{braket}
\usepackage{xcolor}
\usepackage{todonotes}
\definecolor{ugo}{RGB}{158,0,0}

\begin{document}

\title{Optical quantum super-resolution imaging and hypothesis testing
}

\author{Ugo Zanforlin}
\affiliation{Scottish Universities Physics Alliance, Institute of Photonics and Quantum Sciences, School of Engineering and Physical Sciences, Heriot-Watt University, David Brewster Building, Edinburgh EH14 4AS, UK}

\author{Cosmo Lupo}
\affiliation{Dipartimento di Fisica, Politecnico di Bari, 70126 Bari, Italy}
%\affiliation{Department of Physics and Astronomy, The University of Sheffield, Hounsfield Road, S3 7RH Sheffield, United Kingdom}

\author{Peter W. R. Connolly}
\affiliation{Scottish Universities Physics Alliance, Institute of Photonics and Quantum Sciences, School of Engineering and Physical Sciences, Heriot-Watt University, David Brewster Building, Edinburgh EH14 4AS, UK}

\author{Pieter Kok}
\affiliation{Department of Physics and Astronomy, The University of Sheffield, Hounsfield Road, S3 7RH Sheffield, United Kingdom}

\author{Gerald S. Buller}
\affiliation{Scottish Universities Physics Alliance, Institute of Photonics and Quantum Sciences, School of Engineering and Physical Sciences, Heriot-Watt University, David Brewster Building, Edinburgh EH14 4AS, UK}

\author{Zixin Huang}
\email{zixin.huang@mq.edu.au}
\affiliation{Center for Engineered Quantum Systems, Department of Physics and Astronomy, Macquarie University}
\affiliation{Department of Physics and Astronomy, The University of Sheffield, Hounsfield Road, S3 7RH Sheffield, United Kingdom}

\begin{abstract}\noindent

Estimating the angular separation between two incoherent thermal sources is a challenging task for direct imaging, especially when it is smaller than or comparable to the Rayleigh length. In addition, the task of discriminating whether there are one or two sources followed by detecting the faint emission of a secondary source in the proximity of a much brighter one is in itself a severe challenge for direct imaging.
Here, we experimentally demonstrate two tasks for superresolution imaging based on quantum state discrimination and quantum imaging techniques. We show that one can significantly reduce the probability of error for detecting the presence of a weak secondary source, especially when the two sources have small angular separations. In this work, we
reduce the experimental complexity down to a single two-mode interferometer: 
we show that (1) this simple set-up is sufficient for the state discrimination task, and (2) if the two sources are of equal brightness, then this measurement can super-resolve their angular separation, saturating the quantum Cram\'er-Rao bound.
By using a collection baseline of 5.3~mm, we resolve the angular separation of two sources that are placed 15~$\mu$m apart at a distance of 1.0~m with an accuracy of $1.7\%$--this is between 2 to 3 orders of magnitudes more accurate than shot-noise limited direct imaging.

% \textbf{For Nature Comms the abstract structure is motivation and previous work, followed by our work. I moved the first sentence to achieve this. We should also add references, but we could also wait until accepted (PK).}
\end{abstract}
\date{\today}
 
\maketitle

\section{Introduction}
Hypothesis testing, parameter estimation, and imaging are fundamental scientific methods.
Traditionally, the resolution of an imaging system is limited by the Rayleigh criterion \cite{born2013principles,rayleigh1879xxxi}: the minimum angular separation that can be resolved is $\theta_\text{min} \approx 1.22\lambda/D$, where $\lambda$ is the wavelength, and $D$ is the diameter of the lens aperture. Direct imaging of features smaller than $\theta_\text{min}$ is almost impossible since the feature size is as large as its blurring on the image screen. 
Techniques of quantum imaging are often developed with the goal of surpassing the diffraction limit
\cite{brida2010experimental,lugiato2002quantum,shapiro2012physics,Perez12,PhysRevA.77.043809,PhysRevA.79.013827, unternahrer2018super,casacio2021quantum}. For example, ghost imaging \cite{shapiro2012physics,erkmen2010ghost,PhysRevA.77.043809}, quantum lithography \cite{PhysRevLett.87.013602,PhysRevA.63.063407}, and quantum sensing \cite{costa2006use,RevModPhys.89.035002,photonic_quantum_sensing} exploit entangled or correlated sources to enable precision beyond what is achievable classically, while fluorescence super-resolution microscopy \cite{ram2006beyond, thorley2014super, small2014fluorophore,mortensen2010optimized} bypasses the Rayleigh criterion by using engineered sources to suppress shot-noise.

When source engineering is not an option---which is the case for astronomical observations---
quantum techniques can beat the diffraction limit by unlocking all the information about amplitude and phase in the collected light: a recent result for super-resolving a pair of incoherent sources has triggered much interest in the field \cite{PhysRevX.6.031033}, where it was shown that there is no loss of precision associated with estimating the sources' angular separation, even when their separation is smaller than $\theta_\text{min}$.
However, prior to measuring the separation, one needs to ensure that there are two sources and not just one. This task itself becomes difficult when the two sources overlap on the image screen; this is especially true if one source is significantly fainter than the other.
Then, the experimenter is first faced with a hypothesis testing task to determine whether there are one or two sources. 
One straight-forward method would be to use direct imaging (DI) to determine whether a secondary source is present. 
In a diffraction limited system, the image of a point-like object is not a point but has a finite spread characterised by the point-spread function (PSF). If the two sources overlap on the image screen, this blurring presents a severe practical obstacle to direct detection of exoplanets \cite{wright2013exoplanet,fischer2014exoplanet}, especially when one source is much dimmer than the other. 
% For discrimination of two sources with relative intensities $1-\epsilon$ and $\epsilon$ where $\epsilon \ll1$, the probability of error
% %we can quantify this difficulty using the fact that the relative entropy 
% for DI scales as $\epsilon^2$.

Instead, quantum hypothesis testing techniques can be used. Here the task is to determine whether a secondary source exists. 
The goal is to minimise the probability of a false negative (missing the second source).
If we are happy to accept a certain probability of false positives (type-I error), then the probability of a false negative (type-II error), is given by the quantum Stein Lemma \cite{Petz1991,Ogawa2000}. 
This asymmetric error setting is particularly applicable to rare events such as exoplanet identification \cite{wright2013exoplanet,fischer2014exoplanet}, or events with important ramifications such as dimer detection in microscopy~\cite{nan2013single}.
In quantum information theory, the two hypotheses, one source vs two sources, are modelled by two quantum states, $\rho_0$ and $\rho_1$. We consider $n$ incoming photons, and define $\alpha_n$ and $\beta_n$ as the probabilities of type-I and type-II errors, respectively. Given a bounded probability of a type-I error, $\alpha_n < \delta$, the quantum Stein lemma \cite{li2014second,tomamichel2013hierarchy} states that 
\begin{align}
\beta_n = \exp\left[ -n D(\rho_0||\rho_1)   + \sqrt{n b}\,\Phi^{-1}(\delta) + O(\ln n)  \right]
\, ,
\end{align}
where 
\begin{align}
D(\rho_0||\rho_1) = \text{Tr}[\rho_0 (\ln \rho_0 - \rho_1)]
\end{align}
 is the quantum relative entropy (QRE) \cite{wilde2013quantum}, $\Phi(\delta)$ is the Error Function,
%cumulative distribution function for a standard normal random variable:
%$\Phi(y) \equiv  1/\sqrt{2\pi}\int_{-\infty}^y dx \exp(-x^2/2)$
and $b$ is the variance of the QRE \cite{PhysRevLett.119.120501}. 
In the limit of large $n$, the QRE $D(\rho_0||\rho_1)$ dominates $\beta_n$.
The QRE provides a $1/\epsilon$ factor improvement over the classical relative entropy for direct imaging in the error exponent of $\beta_n$ \cite{PhysRevLett.127.130502}, thereby significantly reducing the probability of error,
%for detecting the presence of a weak secondary source, 
even when the two sources have small angular separations.

Once it is established with reasonable confidence that there are two sources, one can use quantum metrology to perform parameter estimation on the angular separation. The precision in the estimation is dictated by the quantum Cram\'er-Rao bound. 
For any density matrix $\rho(\lambda)$ with spectral decomposition \mbox{$\rho(\lambda) = \sum_i p_i \ket{e_i}\bra{e_i}$} that encodes the information of the parameter $\lambda$, the mean square error $\Delta^2\lambda$ is lower bounded by the  quantum Fisher information (QFI) $I_\lambda$,
\begin{align}
\Delta^2 \lambda \geq \frac{1}{n I_\lambda} \, , \qquad
I_\lambda = 2\sum_{i,j}
\frac{ \braket{e_i|\partial_\lambda \rho|e_j} }{p_i + p_j} \, ,
\end{align}
where $n$ is the number of copies of the state.  Unlike in DI where this precision drops to zero when the separation is small compared to the width of the PSF, for two equally bright sources the QFI is finite and independent of the separation. 
The method for obtaining sub-Rayleigh super-resolution through coherent detection of incoherent light is currently an active area of research  \cite{PhysRevLett.117.190802,PhysRevLett.117.190801,PhysRevA.95.063829,PhysRevA.96.063829,PhysRevA.95.063847,PhysRevA.96.062107,PhysRevLett.121.023904,PhysRevLett.117.190801,PhysRevLett.121.180504,PhysRevLett.122.140505,npjQI2019, PhysRevA.99.033847,PhysRevA.99.012305,paur2016achieving,tang2016fault,yang2016far,PhysRevLett.118.070801,PhysRevLett.121.090501,PhysRevLett.121.250503,paur2018tempering,hassett2018sub,zhou2019quantum, Howard19,Pearce2017optimalquantum}. However, the optimal measurement is typically highly nontrivial.
In this paper, we significantly simplify the required experimental complexity: (1) we experimentally demonstrate clear sub-Rayleigh scaling for quantum state discrimination of singular versus binary sources, and (2) we approach the quantum Cram\'er-Rao bound for estimating the angular separation of two sources with equal brightness. Most importantly, the two tasks can be achieved with a single measurement setup: all the above tasks can be performed with a simple two-mode interferometer.

\begin{figure}[hbt]\center
\includegraphics[trim = 0cm 0cm 0cm 0cm, clip, width=1.0\linewidth]{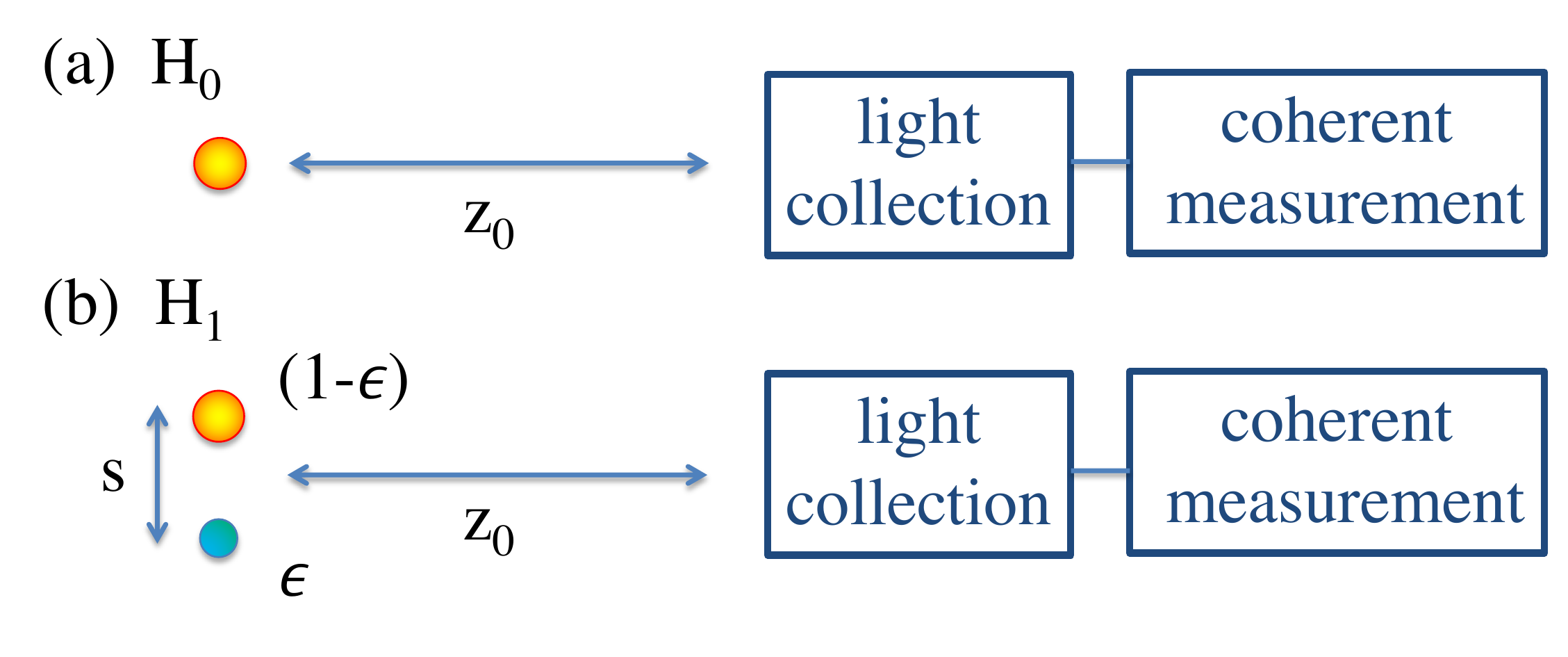} 
  \caption{An optical imaging system is used to discern between two hypotheses, followed by parameter estimation. If hypothesis $H_0$ is true, only is one source is present of intensity $N$; if $H_1$ is true, two sources are present, with total intensity $N$ and the relative intensity is $\epsilon/(1-\epsilon)$. For $H_1$, angular separation between the two sources is $\theta = s/z_0$.
   }\label{f:schematic}
\end{figure}

\section{The Model}

First, consider the task of discriminating between one source or two sources with a separation $s$ 
in the object plane. Hypothesis $H_0$ states that only one source is present, and hypothesis $H_1$ states that two sources are present, centered at $x_0$ with angular separation $\theta = s/z_0$. Furthermore, they have relative intensities $(1-\epsilon)$ and $\epsilon$ respectively. 
We will label a photon originating from the potentially brighter source with intensity $(1-\epsilon)$ as $\ket{\psi_\text{star}}$, and the source with intensity $\epsilon$ as $\ket{\psi_\text{planet}}$. The two states on the image plane are generally non-orthogonal.
The density matrices associated with the two hypotheses $H_0$ and $H_1$ are, respectively
% The states to discriminate between are
\begin{align}
\rho_0 &= \ket{\psi_\text{star}}\bra{\psi_\text{star}} \, , \nn
\rho_1 &= (1-\epsilon)\ket{\psi_\text{star}}\bra{\psi_\text{star}} +
 \epsilon \ket{\psi_\text{planet}}\bra{\psi_\text{planet}} \, .
\end{align}
When imaged using a lens with PSF of size $\sigma$, in the limit that $\theta \leq \sigma, \epsilon \ll 1$ the relative entropy of this state is
\mbox{$D(\rho_0||\rho_1) \approx (\exp[\theta^2/\sigma^2]-1)\epsilon^2/2$} \cite{PhysRevLett.127.130502}. This quadratic scaling in $\epsilon$ formally expresses the challenges of using DI for exoplanet detection, especially when the planet is much dimmer and very close to the star. An almost-optimal quantum measurement, SPADE \cite{PhysRevLett.127.130502}, is able to achieve linear scaling in $\epsilon$ by performing spatial Hermite-Gaussian mode sorting. However, implementing SPADE is experimentally challenging: the setup is sensitive to misalignment \cite{PhysRevX.6.031033}, and the need to split higher order modes is highly non-trivial.

If instead of a lens we place two optical collectors, $d_1$ and $d_2$, separated by $d = |d_1 - d_2|$, and at a distance $z_0$ from the sources (Fig.~\ref{f:schematic}), then the states $\ket{\psi_\text{star}}$ and $\ket{\psi_\text{planet}}$ can be described as:
\begin{align}
\ket{\psi_\text{star}} = \frac{1}{\sqrt2}(\ket{d_1} + e^{i \phi}\ket{d_2}) \, , \nn
\ket{\psi_\text{planet}} = \frac{1}{\sqrt2}(\ket{d_1} + e^{i \psi}\ket{d_2})  \, ,
\end{align} 
where $\phi,\psi$ are the optical path differences of the sources to the two collectors. In the paraxial regime, these are
\begin{align} 
\phi 
% &= k z_0 \left( \frac{(x_0 + s/2 - u_2)^2}{2 z_0^2} - \frac{(x_0 + s/2 -  u_1)^2}{2z_0^2} \right)^{1/2} \nn
       \approx  \frac{k d \theta}{2} \, , \qquad
\psi  \approx \frac{-k d \theta}{2} \, .
\end{align}
In the limit of $\epsilon \ll1$, the relative entropy between $\rho_0$ and $\rho_1$ is approximately
(see Supplemental Material)
\begin{align}\label{eq:qre_approx}
D(\rho_0||\rho_1) \approx \frac{\theta ^2 k^2 \epsilon  d^2}{4 } \, .
\end{align}
% \textbf{Is this the QRE? Should we write $D$ instead of $S$?}
% 
Equation~\eqref{eq:qre_approx} is also linear in $\epsilon$, thus has a factor $1/\epsilon$ improvement compared to the classical counterpart; an optimal measurement that saturates the QRE is by placing a phase shifter and a 50:50 BS after the two collectors, followed by photon counting. Given an imperfect interferometer with visibility $\nu$, if there is no planet ($H_0$ is true), then the probabilities that the photon is detected at detectors $a$ or $b$ are
\begin{align}
p_{H_0}(a) = \frac{1}{2}(1+\nu\cos(\psi_1 + \alpha)) \, , \nn
p_{H_0}(b) = \frac{1}{2}(1-\nu\cos(\psi_1 + \alpha)) \, .
\end{align}

\begin{figure}[t]\center
\includegraphics[trim = 0cm 0cm 0cm 0cm, clip, width=1.0\linewidth]{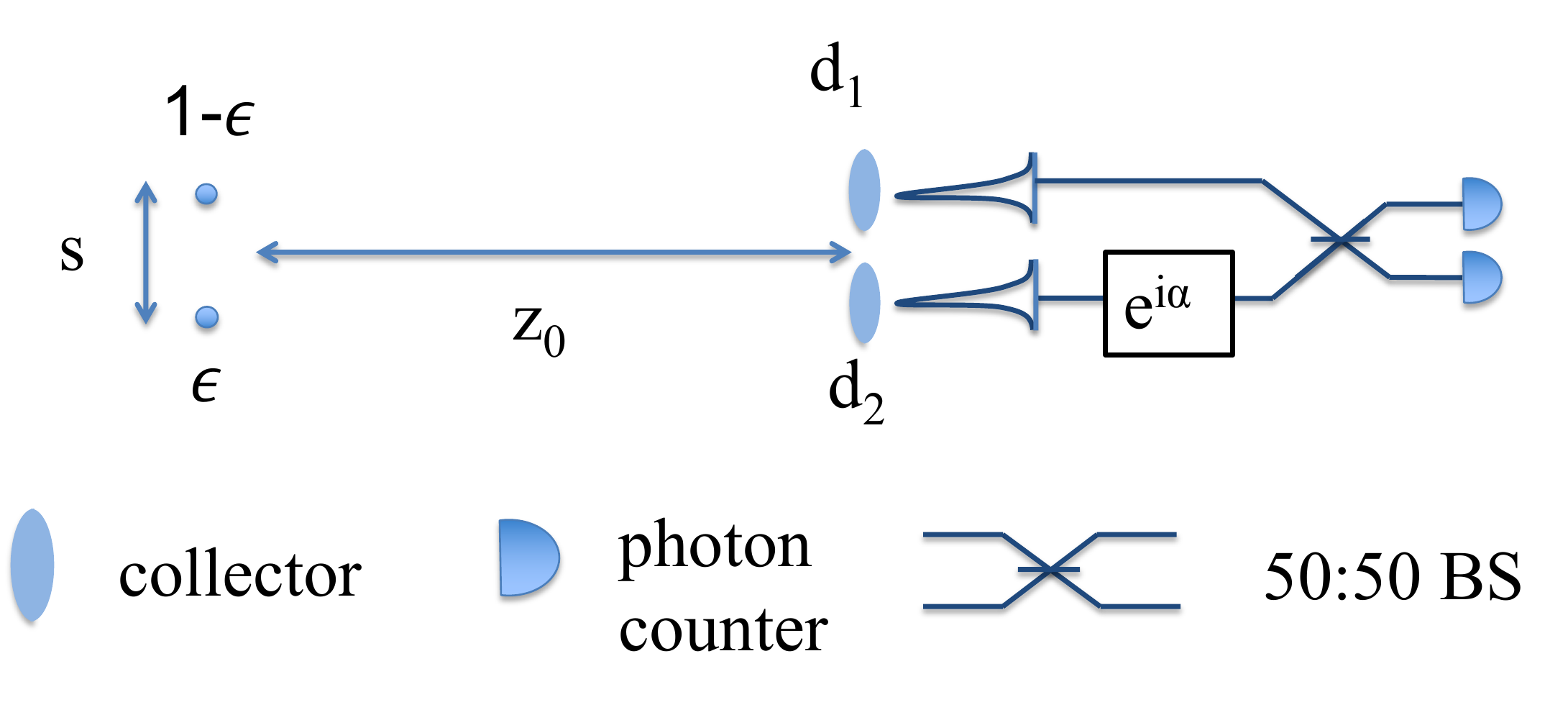}
\caption{ Schematic of two sources with a separation of $s$ in the object plane, at a distance $z_0$ from the collectors. Two collectors at $d_1$ and $d_2$ direct light into a two-mode interferometer consisting of a phase shift of $\alpha $  and a 50:50 beam splitter, followed by photon counters.\label{f:schematic}}
\end{figure}

\begin{figure*}[t!]
\centering
\includegraphics[width=\linewidth]{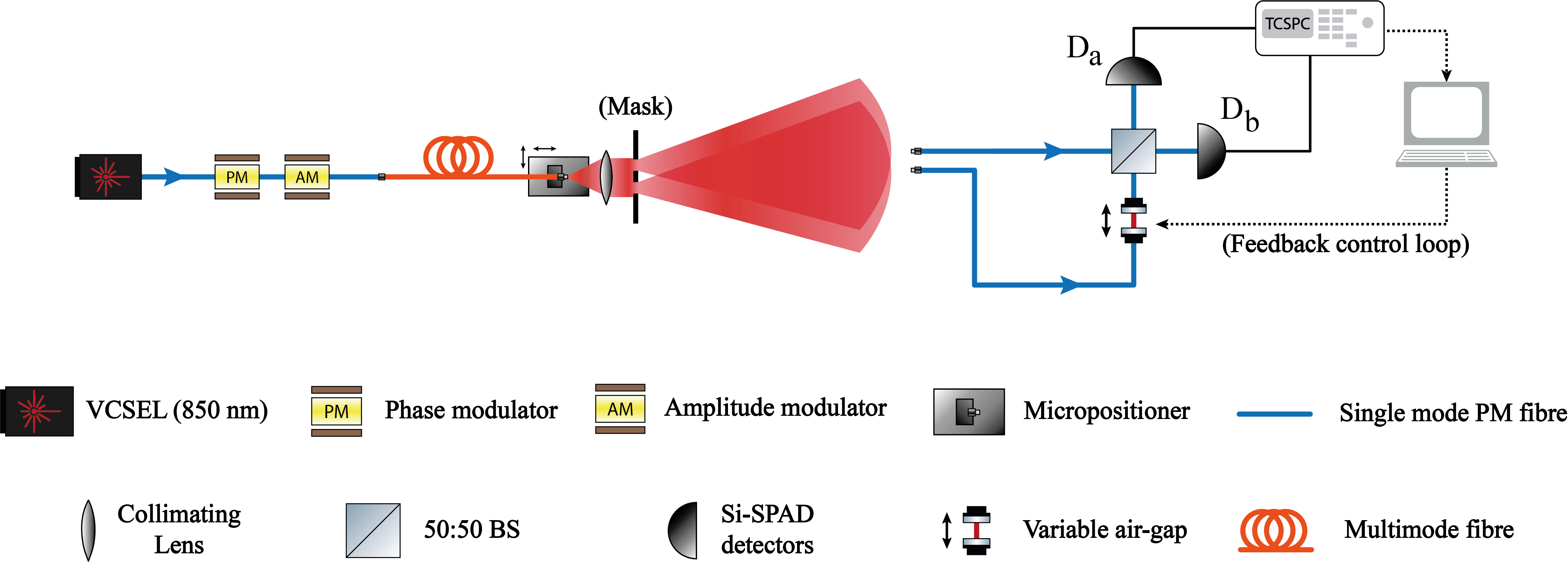}
\caption{Experimental set-up. A VCSEL operated in pulsed mode generates coherent states that are phase and amplitude modulated to reproduce a pseudo thermal state. These states are then coupled into a multimode fibre and then collimated to a custom optical mask shaping the light beam into two pseudo-point-like sources. At 1~m distance, two single-mode polarization-maintaining fibers collect the transmitted beam and perform mode sorting via a balanced interferometer. Two Si-SPADs register photon detection events and store their information onto a PC for post-processing via a TCSPC module. A feedback control system is used for interferometric stabilization via an adjustable air gap.}
\label{f:exp_setup}
\end{figure*}

\noindent
Otherwise, if $H_1$ is true, then the probabilities that the photon is detected are
\begin{align}
p_{H_1}(a) = \frac{1}{2}[&(1-\epsilon)(1+\nu\cos(\psi_1 + \alpha)) + \nn
  &\epsilon(1+\nu\cos(\psi_2 + \alpha))] 
  \, , \nn
p_{H_1}(b) = \frac{1}{2}[&(1-\epsilon)(1-\nu\cos(\psi_1 + \alpha)) + \nn
&\epsilon(1-\nu\cos(\psi_2 + \alpha)) ]
\, .
\end{align}
The classical relative entropy (CRE) for this measurement is maximised for 
\begin{align}
\alpha \approx k d [ \epsilon(  x_0 + s)/z_0 + (1-\epsilon)(x_0)/z0] \, ,
\end{align}
and matches the QRE. Intuitively, this corresponds to the the point where 
$|p_{H_1}(a) - p_{H_0}(a)|=|p_{H_1}(b) - p_{H_0}(b)|$
is maximised. 
%\textbf{Which $p(a)$ are we talking about here, $H_0$ or $H_1$?}

When the source intensities are equal, the QFI for the above state is
\begin{align}
I_\theta = \frac{k^2}{4} (d_1 - d_2)^2,
\end{align}
which is constant in the effective pupil size $|d_1 - d_2|$ and independent of the separation. Hence, the separation can be estimated well beyond the diffraction limit.

The measurement that achieves the maximum relative entropy is the same as the one that saturates the quantum Cram\'er-Rao bound dictated by the QFI.
Define $a^{\dagger'}_{d_1}(a^{\dagger'}_{d_2})$ to be the creation operator at the collector position $d_1~(d_2)$. 
The phase and the beam splitter transform the operators as
\begin{align}
    a^{\dagger'} _{d_1} \rightarrow\frac{1}{\sqrt2}(a^\dagger_{d_1} + a^\dagger_{d_2}) \, , \nn
        a^{\dagger'}_{d_2} \rightarrow\frac{e^{i \alpha}}{\sqrt2}(a^\dagger_{d_1} - a^\dagger_{d_2}) \, ,
\end{align}
The probabilities of detecting the photon at either detector are 
\begin{align}\label{eq:equal_intensity}
p_a(\phi,\alpha,\nu) = \frac{1}{2}\left( 1+ \nu \cos(\alpha)\cos\left[ \phi \right] \right) \, ,\nn
p_b(\phi,\alpha,\nu) = \frac{1}{2}\left( 1- \nu \cos(\alpha)\cos\left[ \phi \right] \right) \, .
\end{align}
The maximum classical relative entropy and Fisher information are achieved around the phase values $\alpha = 0, \pi$, which coincide with the QRE (see Supplemental Material) and QFI \cite{PhysRevLett.124.080503} respectively.

\section{Experimental set-up}
The experimental set-up is depicted in Fig.~\ref{f:exp_setup}. A fiber-coupled vertical cavity surface-emitting laser (VCSEL) with 848.2~nm central wavelength (0.11~nm FWHM) is operated in pulsed mode at a repetition rate of 1~MHz. This specific wavelength is chosen as it provides a good trade-off between single-photon detection efficiency ($\approx 40\%$) with commercially available thick-junction silicon single photon avalanche diodes (Si-SPADs) detectors and tolerable optical loss in silica fibers ($\approx 2.2$~dB/km) \cite{buller_2009}. The resulting coherent states are then coupled into two electro-optic modulators (EOMs) enabling phase and amplitude modulations of the individual coherent states. An external arbitrary waveform generator (AWG) electrically drives the two modulators by means of randomised modulation patterns so that the resulting optical states resemble a pseudo-thermal source, required for the incoherent sources specified by the model and tested using a Hanbury Brown and Twiss interferometer. This modulation approach provides absolute control over each coherent state emitted by the source, including preserving the coherent state for use in interferometric measurements. In the results presented in this paper, we alternate the pseudo-thermal state with a coherent state that acts as a reference. The reference pulses provide the necessary interferometric stabilisation that is controlled via feed back from the two detectors, after the two pulses interfere at the beamsplitter. The alternate set of thermally modulated states can then undergo the mode sorting technique detailed in our model (see Supplemental Material).

After the phase and amplitude modulation, the pseudo-thermal states are then coupled into multimode optical fibres (8~m in length) in order to maximize mode dispersion and reduce wavefront spatial correlations due to the initial coupling of the VCSEL to single mode based optical components. The final thermal radiation is coupled into an adjustable aspheric collimator lens providing precise alignment with the remaining free-space optical components. Two pseudo-thermal sources are extracted from the collimated beam via a custom-made optical mask with two circular pinholes etched onto the surface, effectively reproducing two idealized point-like sources corresponding to the two distant stars of our model. Different etched patterns were fabricated using laser-written lithography in order to study a wide range of configurations with pinhole dimensions ranging from 10 to 50~$\mu$m in diameter and with spatial separation spanning from just 15~$\mu$m to almost 1~cm (see Supplemental Material for further details).

A neutral density filter is mounted on a separate movable micro-positioner block (not shown) placed in front of one of the two pinholes reducing the transmitted optical power through one of the pinholes. This configuration creates a controlled intensity imbalance between the two pseudo thermal sources effectively creating one bright source (a distant star) and one dimmer source (a distant exoplanet). At 1~m from the mask, two single mode polarization maintaining optical fibers, separated by 5.3~mm, are mounted on a micropositioner block (not shown) coupling the transmitted light beams into a balanced interferometer whose output modes are monitored by Si-SPAD detectors. An adjustable air-gap is placed in one of the two optical paths allowing us to loss-balance the interferometer as well as providing direct control over the optical path-length difference. A time-correlated single photon counting unit processes the generated timetags with 1~ps resolution enabling fast readout times as well as full digital post-processing. For each configuration of the set-up, 25 individual measurements are taken with a 5~s integration time in order to reduce Poissonian errors associated with photon-count data. An active feedback mechanism is implemented to ensure high interferometric visibility ($> 99\%$) during the entire duration of the data acquisition.

\section{Results}

\begin{figure}[h!]\center
\includegraphics[trim = 0cm 0cm 0cm 0cm, clip, width=1.0\linewidth]{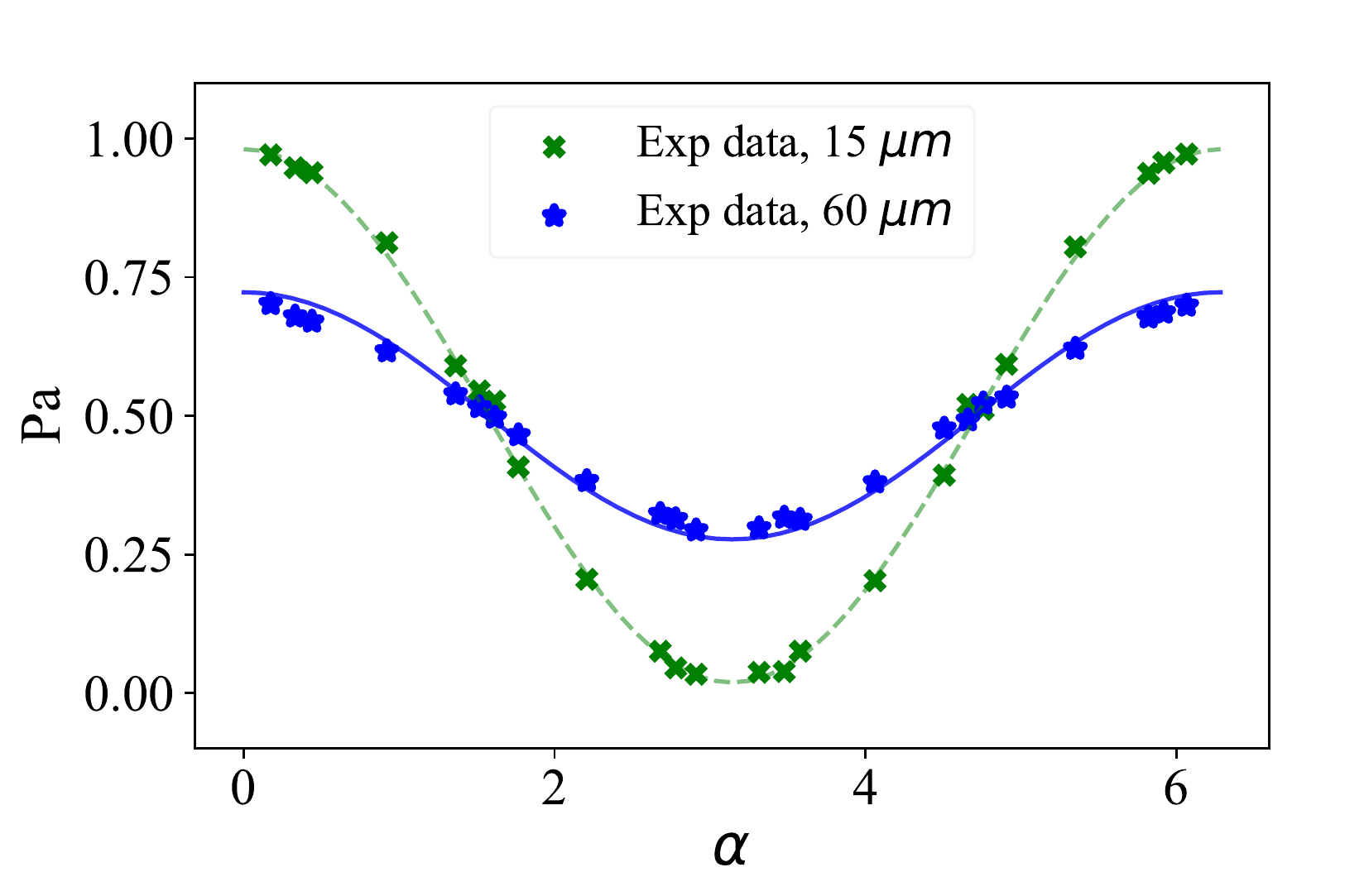} 
  \caption{The probability of the photon output at one of the detectors, as a function of the applied phase $\alpha$, which is adjusted using the distance of the air-gap. The data shown are for $\epsilon = 0.5$, for physical separations of $1.5 \times 10^{-6}$ and $60\times 10^{-6}$; the angular separation are 1.48 $\times 10^{-6}$ and $59\times 10^{-6}$ respectively. 
 }\label{f:qre0}
\end{figure}

% \textbf{I think we need to show one more graph before the actual results}

\begin{figure}[t]
\includegraphics[trim = 0cm 0cm 0cm 0cm, clip, width=1.0\linewidth]{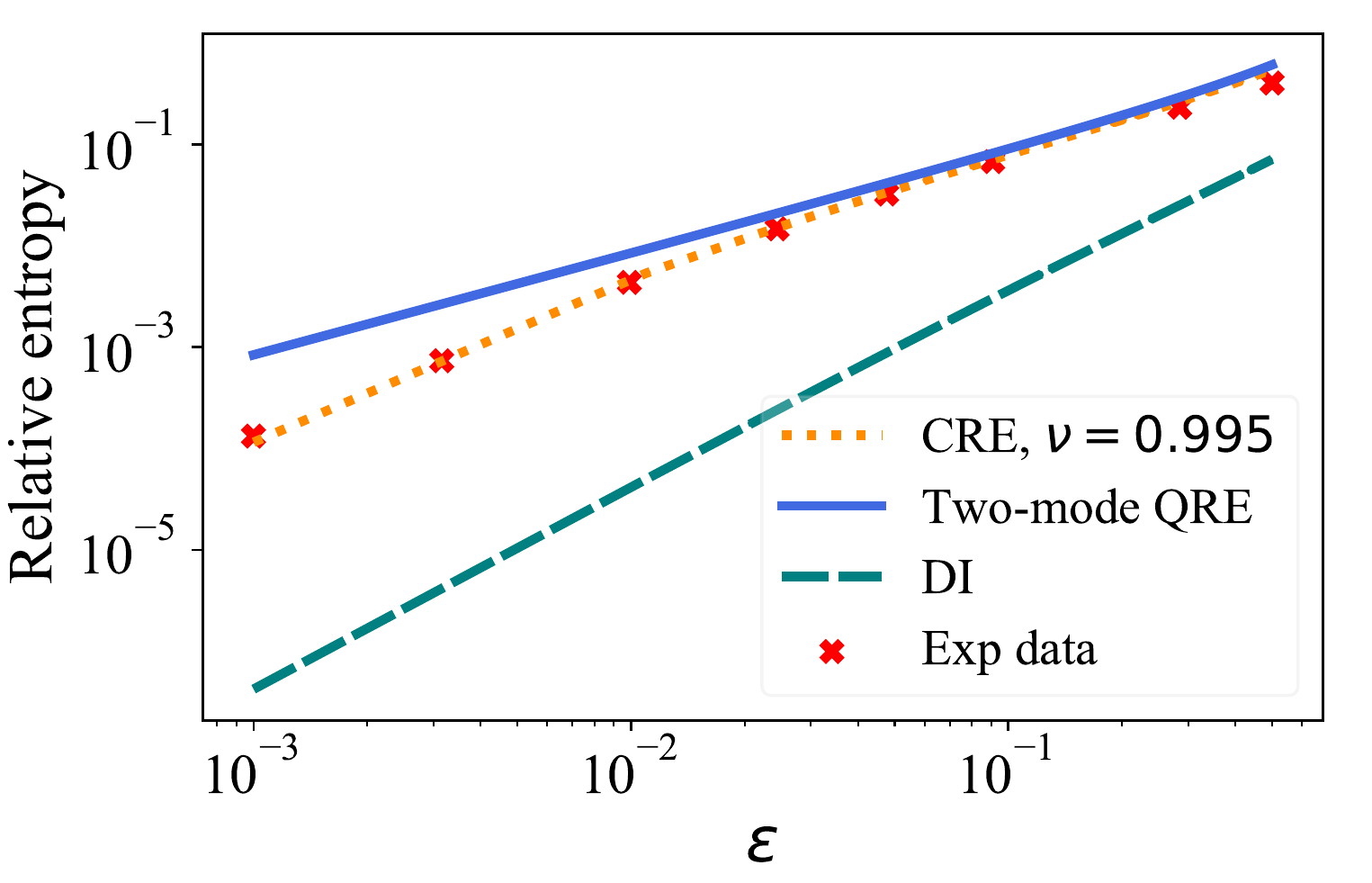} 
\caption{Relative entropy of the two hypothesis for different values of $\epsilon$. The plots shows: (1) the QRE of the two-mode state (blue solid line), (2) the CRE of the measurement maximised over $\alpha$, given $\nu = 0.995$ (orange dotted line), (3) the CRE for shot-noise limited direct imaging (DI, teal dashed line), and (4) the experimental data points (red crosses). }\label{f:qre}
\end{figure}

%\textbf{(this looks like 12.2 arc seconds, is that right? Maybe use this unit, as it is more common? Hubble has a resolution of 0.05 arc seconds with a mirror of 2.4~m, so we're twice as good as Hubble)}
We experimentally measured the probability of the photon arriving at detectors $a$ and $b$, and computed the relative entropies. As an example, in Fig.~\ref{f:qre0} we show the probability of the photon arriving at detector $a$ for $\epsilon = 0.5$ and angular separations of
 $1.5\times 10^{-5}$~rad and $5.9\times 10^{-5}$ ~rad. In Fig.~\ref{f:qre}
we present the CRE of the measurement for different values of $\epsilon$ using an angular separation of $ 5.9\times 10^{-5}$ rad. 
For comparison, we also show the relative entropy for direct imaging using a lens of the same diameter (5.3~mm, assuming a Gaussian PSF).
Fig.~\ref{f:qre} shows the distinct difference in scaling in $\epsilon$ between our method and DI. For $\epsilon > 10^{-2}$, we see that the two-mode CRE
matches the QRE well. Due to experimental imperfections, around $\epsilon \sim 10^{-3}$ the achievable relative entropy has significantly deviated from the ideal quantum case, but still surpasses the DI limit by two orders of magnitude.

% For the QRE is $1.9 \times 10^{-4}$ and $0.58$ bits for 
% $\epsilon \approx 0.001$ and $\epsilon = 0.5$ respectively, which means that which means that the type-II error is negligible for our case after 60000 photons are used.\textcolor{red}{the point at $\epsilon=0.001$, QRE is 
% 0.00019197 bits (QRE in natural bits divided by log(2)}

We use maximum likelihood estimation to first extract the optical path difference $\phi$ between the source and the two collectors, and then
obtain an estimator for the angular separation $\theta$. Our method for extracting $\phi$ is a simpler version of the phase estimation method used in Refs.~\cite{PhysRevLett.85.5098, PhysRevA.90.023856,PhysRevA.95.053837}.
We can determine $\phi$ directly from the detection statistics. 
%To choose the estimate, it is useful to determine a probability density function for $\phi$ based on the detection results.
After $m$ detection events, the vector of measurement outcomes is $\vec \mu_m = (\mu_1,\mu_2,\ldots,\mu_m)$, where each element $\mu_j\in\{a,b\}$ corresponds to the detector $a$ or $b$ that signalled the presence of the photon. The probability density function for $\phi$ follows from Bayes' theorem, and is given by \cite{PhysRevLett.85.5098, PhysRevA.90.023856,PhysRevA.95.053837} 
\begin{align}
\label{eq:tot}
 P(\phi|\vec \mu_{m}, \alpha,\nu)&\propto P(\mu_m|\phi, \alpha,\nu)\mathcal{P}(\phi|\vec \mu_{m-1},  \alpha,\nu) \, ,
\end{align}
% When the adjustable phase $\alpha$ is constant this reduces to
% \begin{align}
% P(\phi) &\propto p_a^{m-l}(\phi|\alpha,\nu) p_b^{l}(\phi|\alpha,\nu)\, ,
% \end{align}
where the proportionality constant is determined by normalising the distribution.

% \noindent where $N$ is the total number of photons detected.

\begin{figure}[t]
\includegraphics[trim = 0cm 0cm 0cm 0cm, clip, width=1.0\linewidth]{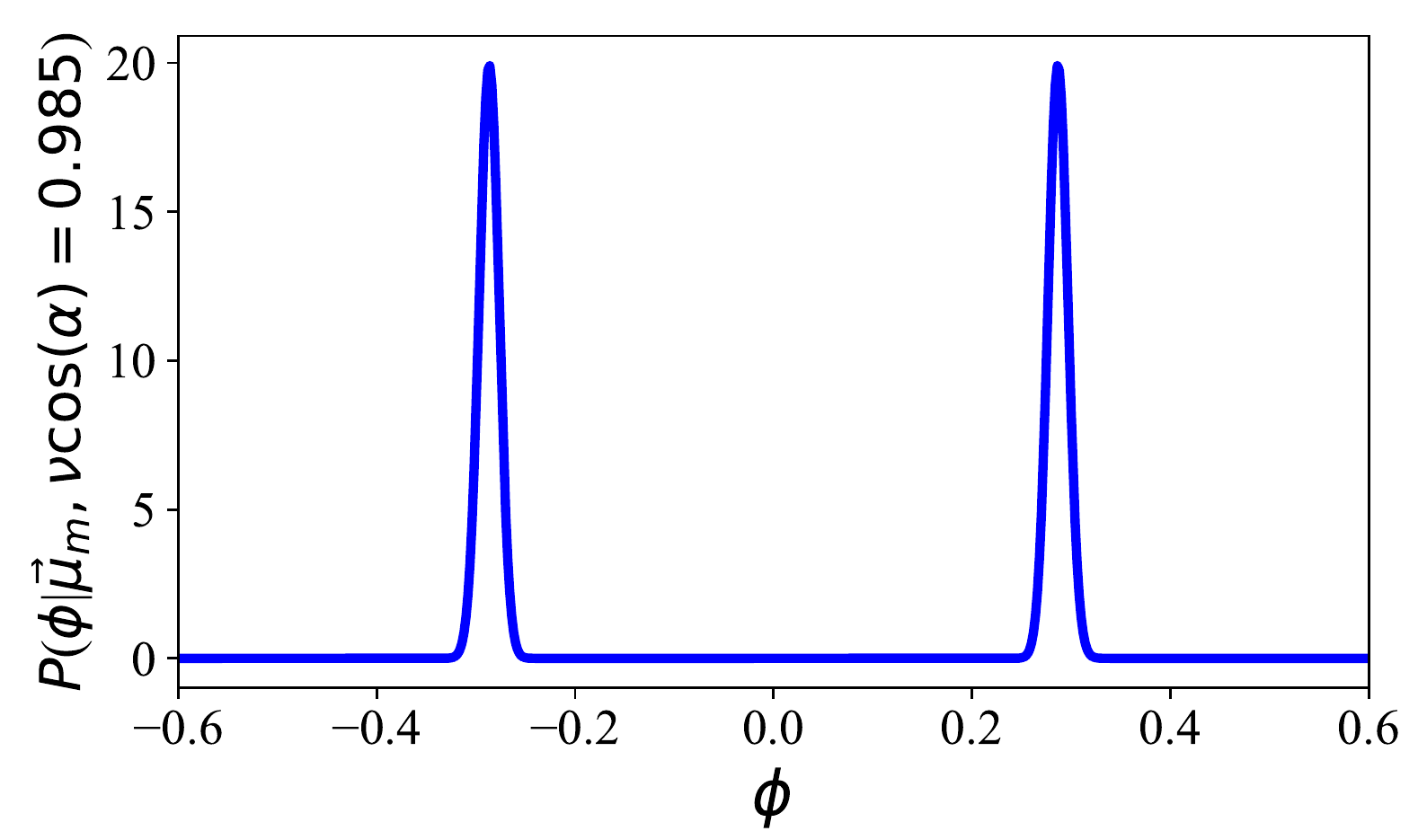} 
  \caption{The probability density function for $\phi$, after 12740 detection events which 1478 were from detector $b$.}\label{f:pdf}
\end{figure}

Prior to any detected photons, we assume no knowledge of $\phi$, and the corresponding prior distribution is therefore $\mathcal{P}_0(\phi) = 1/(2\pi)$. In order to obtain an analytic form for $P(\phi|\vec \mu_m, \alpha,\nu)$, we express it as a Fourier series
\begin{align}\label{eq:fourier}
P(\phi|\vec \mu_m, \alpha,\nu) = \frac{1}{2\pi} \sum_{k = -m} ^m a_k e^{i k \phi}\, ,
\end{align}
where $a_k$ depends on $\vec \mu_m $, $\alpha$ and $\nu$. After each detection event, we can write the updated distribution in this Fourier form as well. 
For example, if detector $b$ fires, then
\begin{align} \label{eq:update}
\mathcal{P}(\mu=b|\phi,\alpha,\nu) = \frac{1}{2}[1 - \nu \cos (\alpha) \cos(\phi) ] \, .
\end{align}
which we can rewrite as
\begin{align}
\mathcal{P}(\mu=b|\phi,\alpha,\nu) = \frac{1}{2} - \frac{1}{4} \nu \cos \alpha  e^{i \phi}  - \frac{1}{4} \nu \cos \alpha  e^{- i \phi} \, .
\end{align}
Therefore the update coefficients are $a_0 =\pi, a_1 = a_{-1} = - \frac{\pi}{2} \nu \cos (\alpha)$. 
The factor $\nu \cos (\alpha)$ is computed directly from the coherent state statistics (see Supplemental Material for details). Before the first detection (the prior distribution), Eq.~\eqref{eq:tot} contains only one term, $a_0=1$. After each detection event the number of Fourier coefficients grow by 2 (the $\pm m$ terms in the Fourier expansion). The coefficients $a_k$ are updated using Eqs.~\eqref{eq:equal_intensity} and \eqref{eq:tot}. As an example, Figure~\ref{f:pdf} shows the probability density function $P(\phi)$ after 12\,740 detection events where $1478$ were output at detector $b$, with $\nu \cos(\alpha) = 0.981$. Since $\cos(\phi)$ is an even function, there are two peaks, symmetrically placed around zero. We require only the magnitude of $\phi$ in the estimation of the angular separation $\theta$.

Following maximum likelihood estimation, the value of $\phi$ at the maximum of $P(\phi)$ becomes our estimate, and the estimate of the separation $\theta$ is then given by
\begin{align}
\hat \theta_\text{est} = 2 |\phi|/(kd)\, .
\end{align}
Once this estimate is obtained, we use the mean-square error (MSE) to quantify the precision, given by
\begin{align}
\text{MSE}(\theta) = \Delta^2\theta + (\bar \theta - \theta_\text{true})^2\, .
\end{align}
Here $\bar \theta$ is the mean value of the estimates, and $\theta_\text{true}$ is the true value of the angle, which in this case is accessible via direct measurement. The MSE is equal to the variance for unbiased measurements and appropriately penalizes biased estimates as well.

\begin{figure}[t]
\includegraphics[trim = 0cm 0cm 0.0cm 0cm, clip, width=1.0\linewidth]{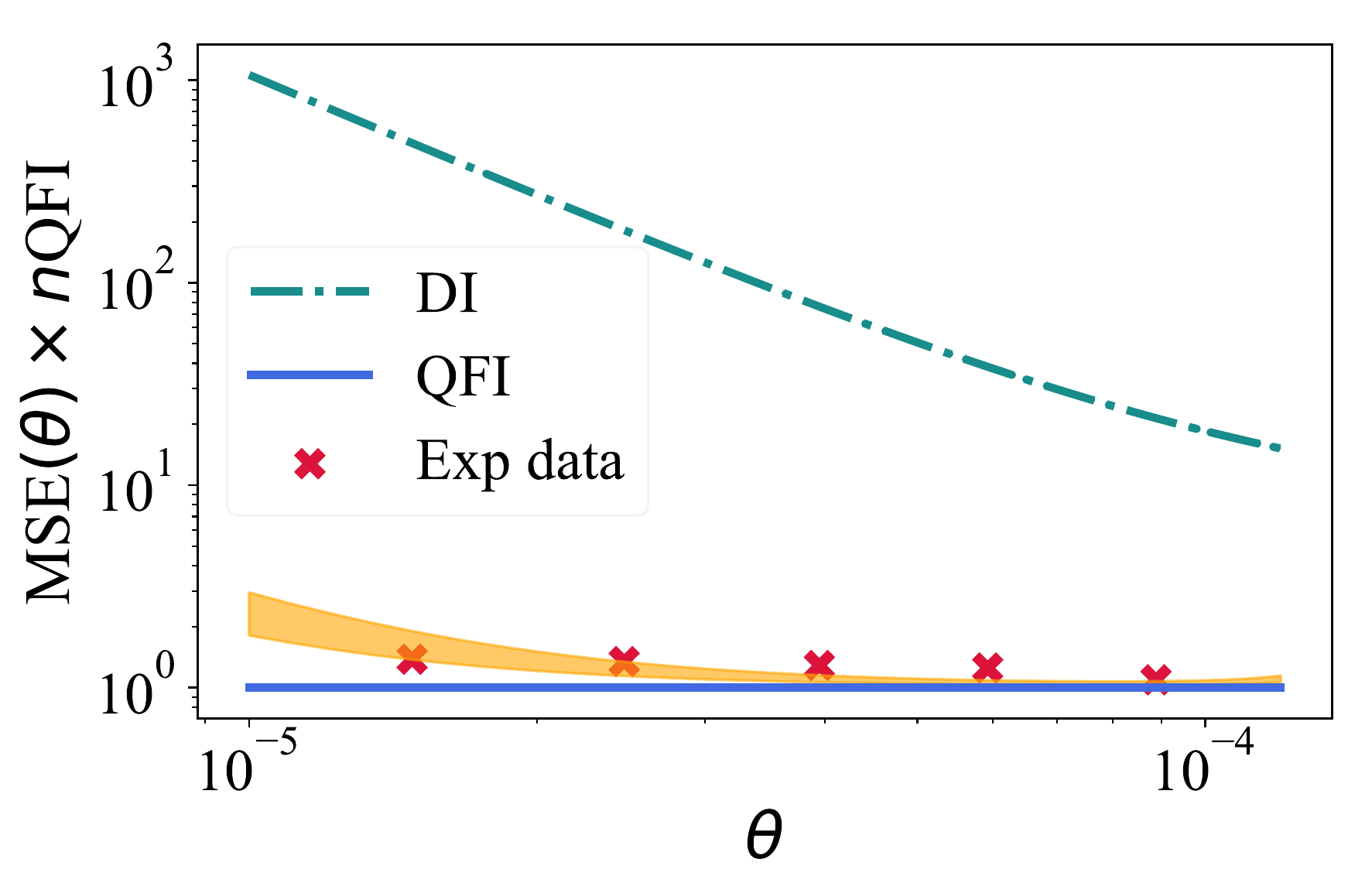} 
  \caption{ The mean squared error (MSE) of estimating the angular separation between two equally bright sources, normalised by the QFI, for different values of angular separation. The quantum Cram\'er-Rao bound, which equals to 1 on this plot (blue solid line). The plot shows (1) the MSE for shot-noise limited DI (teal dotted-dashed line); (2) the Fisher information for an interferometer with a visibility-phase factor $\nu \cos(\alpha)$ between $0.965$ and $0.985$ (orange shaded region); (3) the experimentally achieved MSE (red crosses).}\label{f:fi}
\end{figure}

For each value of the angular separation we obtained 25 different estimates, each detecting approximately $n\approx 60\,000$ photons. Figure~\ref{f:fi} shows
the MSE multiplied by $n\times I(\theta)$.
The experimental data points are indicated by red crosses, and the achievable precision for shot-noise limited DI (using a circular lens of diameter 5.3~mm) is indicated by the dash-dotted line. 

Experimentally, the data was collected with the factor $\nu\cos(\alpha)$ tuned to between 0.96 and 0.985. This is the shaded orange region in Fig.~\ref{f:fi}. The quantum Cram\'er-Rao bound is equal to 1 in this figure (blue solid line). We obtained unbiased estimates for values of angular separation $\theta$ that dramatically violate the Rayleigh limit. When $\theta = 15\times 10^{-6}$~rad, the root-mean-square errors are within $1.7\%$ of the real value, which is two to three orders of magnitude more accurate than what is achievable with DI using a lens of the same diameter. For all the measured angular separations the MSE stayed within a factor 2 of the quantum Cram\'er-Rao bound.

\vspace{3mm}
\section{Conclusions}

In this work we have analysed theoretically, and experimentally demonstrated, two tasks for super-resolution imaging based on quantum state discrimination and quantum parameter estimation.
Estimating the angular separation between two sources is a challenging task for direct imaging, especially when their angular separation is smaller than the point spread function of the imaging system. The task of determining whether there are one or two sources is in itself a difficult task, especially when one source is much dimmer than the other. Until now, the (almost) optimal measurement methods are experimentally challenging.
In this work, we solved both these problems, and have reduced the experimental complexity down to a simple two-mode interferometer: 
we show that a simple set-up achieves sub-Rayleigh scaling for the state discrimination task, and if the two sources are of equal brightness, then this measurement can optimally estimate their angular separation, saturating the quantum Cram\'er-Rao bound.

Our experiment also shows a practical optical setup that could potentially be integrated with current stellar interferometers, however, this would require a different approach for the stabilisation of the interferometer. For example, the stabilisation could be provided by a ground-based coherent source or an artificial guide star which are suitably multiplexed into the interferometry system.
Future work could explore the hypothesis testing for discriminating between multiple sources of different brightness, composite hypothesis testing, and the number of modes the interferometer would require for such tasks.

\section{Acknowledgements}
This work was supported by the UK Engineering and Physical Sciences Research Council projects EP/T001011/1; EP/T00097X/1; EP/S026428/1; EP/M006514/1. This work is funded in part by the EPSRC grant Large Baseline Quantum-Enhanced Imaging Networks, Grant No.~EP/V021303/1.
ZH is supported by a Sydney Quantum Academy Postdoctoral Fellowship.

\bibliography{qre_qfi}

\clearpage
\widetext
% \appendix
\begin{center}
\textbf{\large Supplemental Materials}
\end{center}

\setcounter{equation}{0}
\setcounter{figure}{0}
\setcounter{table}{0}
\makeatletter
\renewcommand{\theequation}{S\arabic{equation}}
\renewcommand{\thefigure}{S\arabic{figure}}
\renewcommand{\bibnumfmt}[1]{[S#1]}
\renewcommand{\citenumfont}[1]{S#1}

\section{Pseudo thermal source generation}
\begin{figure}[h!]
\centering
\includegraphics[width=0.35\linewidth]{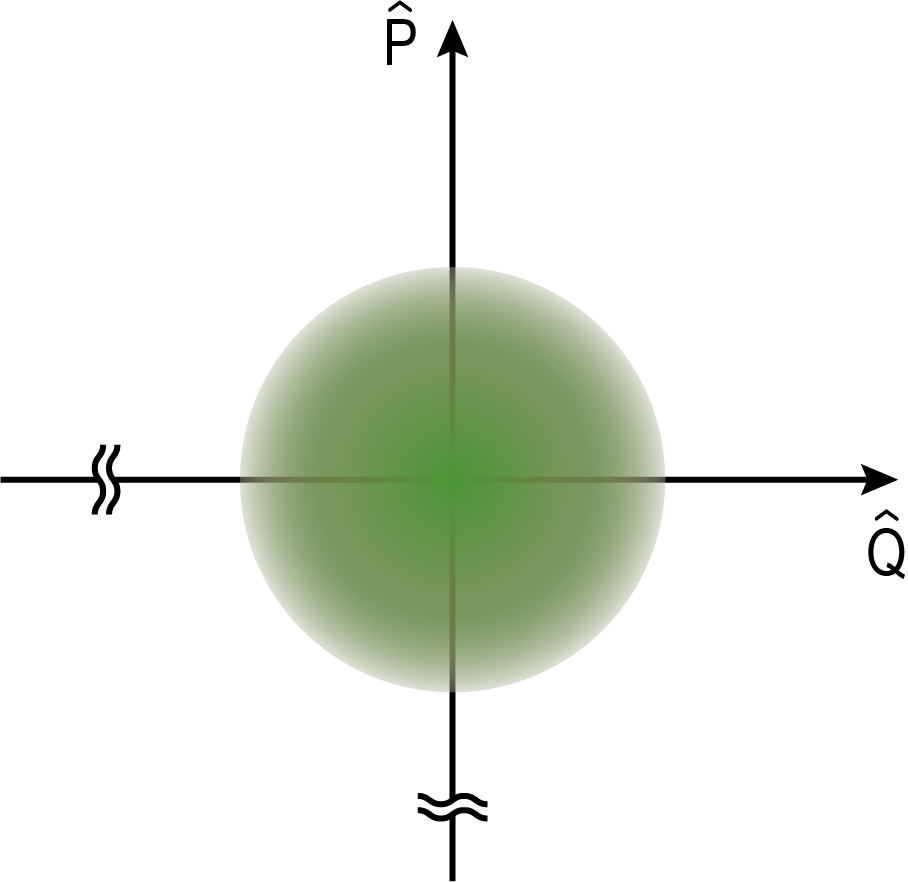}
\caption{Phasor diagram representation of a thermal state. This state has a well defined mean photon number but undefined phase. $\widehat{\text{P}}$ and $\widehat{\text{Q}}$ are the momentum and position operators respectively associated with a quantum harmonic oscillator framework.}
\label{f:phasor_diag_th}
\end{figure}

Thermal radiation is a semi-classical form of radiation characterised by a well defined optical intensity but undefined phase \cite{loudon2000}. Its representation on a phasor diagram is that of a symmetric blurred circle centred around the axes' origin (see Fig.~\ref{f:phasor_diag_th}). Thermal states are generally associated with optical radiation with a reduced temporal and spatial coherence \cite{Mehta2010, Ranganath2008} which limit their use for interferometric measurements. However, our work required precise control over the interferometer's reference phase which could not be achieved by solely implementing a thermal source like an LED. Thankfully, the semi-classical nature of thermal states allows us to express their mathematical representation as a collection of individual coherent states weighted by a suitable quasi-probability distribution, i.e. the Glauber-Sudarshan \mbox{P-function} \cite{Glauber1963}:

\begin{equation}
\rho_{th} = \int P(\alpha) \ket{\alpha}\bra{\alpha} d^2{\alpha}
\end{equation}
where $P(\alpha)$ is the normalized P-function.

\begin{figure}
\centering
\includegraphics[width=0.6\linewidth]{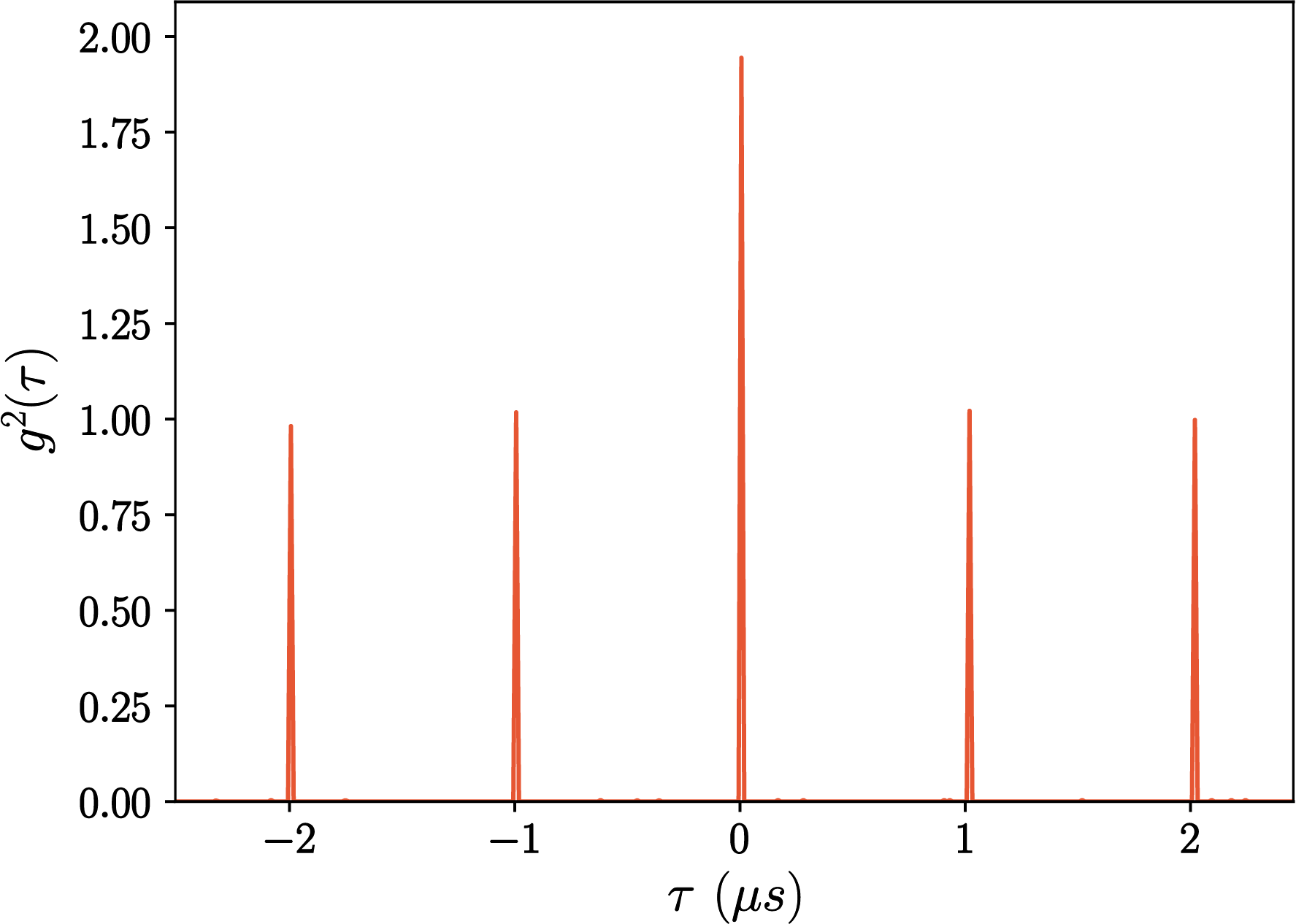}
\caption{Experimental $g^{(2)}(\tau)$ autocorrelation function of the pseudo thermal source. The central peak at zero delay time shows a $g^{(2)}(0) = 1.977 \pm 0.003$ which is in close agreement with a theoretical value of a true thermal state. The coincidence peaks are separated by 1~$\mu$s which is consistent with the clock repetition rate of 1~MHz. All results have been collected in real-time via the QuCoa (PicoQuant) software with a total integration time of 60~s in order to limit evaluation errors.}
\label{f:g2_exp}
\end{figure}

Referencing Fig.~3 in the main manuscript, the phase and amplitude modulations required to reproduce the correct \mbox{P-function} of a thermal state were provided via the two electro-optic modulators (EOMs) fibre-coupled to the laser source (a vertical cavity surface-emitting laser (VCSEL)). A dual-channel arbitrary waveform generator provided independent electrical driving voltages to the EOMs which in turn applied a variable phase and amplitude modulation onto the individual coherent states generated by the VCSEL. Both modulators were LiNbO\textsubscript{3} based with low insertion and coupling loss ($\approx 0.25$~dB) and low DC control voltages ($\approx 2.5$~V for phase inversion). The amplitude modulators comprised two laser inscribed waveguides configured in a Mach-Zehnder pattern where an external RF signal would change locally the refractive index of one arm actively altering the output power of the device. In order to reduce any external interference during their operation, the EOMs were mechanically and thermally isolated from the environment enhancing their operational stability. The final pseudo thermal source was tested by means of a Hanbury Brown and Twiss like experiment \cite{brown1956} were second order correlations between the two detectors monitoring the interferometer were measured for different time delays via the Qucoa analysis software for the HydraHarp 400 (PicoQuant) TCSPC module. Fig.~\ref{f:g2_exp} shows the computed $g^{(2)}(\tau)$ as a function of the time delay between detectors D\textsubscript{A} and D\textsubscript{B} for a pulsed source with a repetition rate of 1~MHz. The results showed a maximum at zero delay of $g^{(2)}(0) = 1.977 \pm 0.003$, in close agreement with the theoretical value of a true thermal source.

\section{Interferometric calibration}
\begin{figure}
\centering
\includegraphics[width=0.9\linewidth]{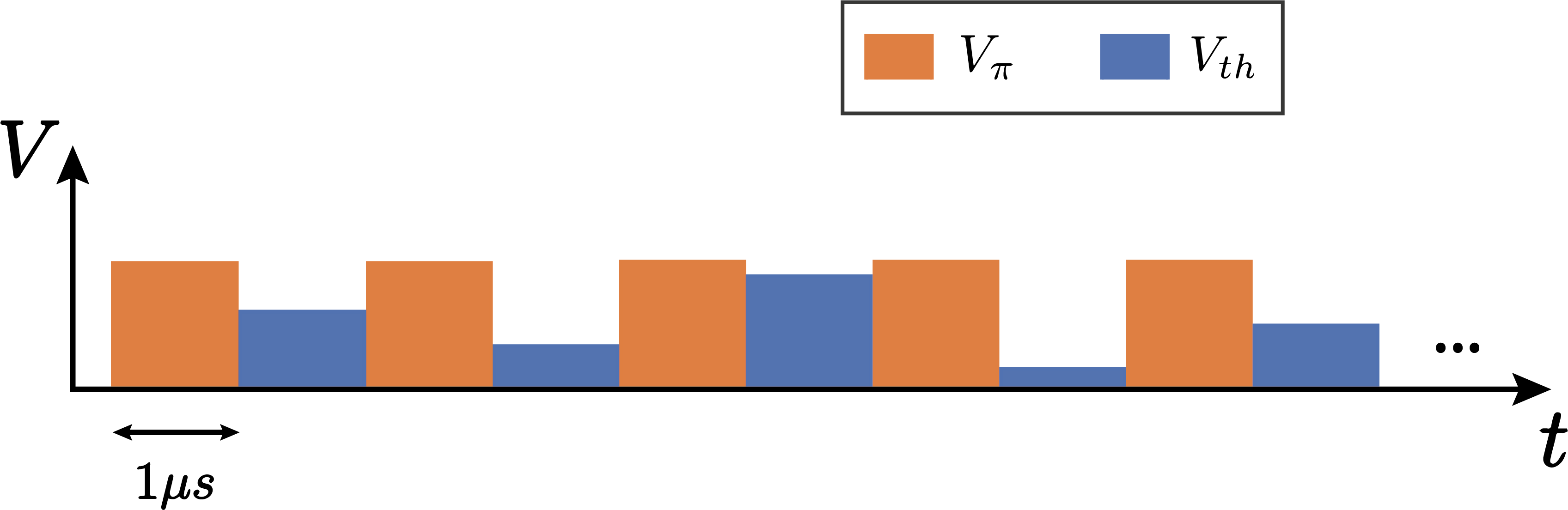}
\caption{Diagram of a one-shot phase modulation signal to the EOM. The multiplexed reference signal is repeated every two electrical pulses (orange rectangles) effectively halving the clock repetition rate to 500~KHz. The thermal  signal instead (blue rectangles) is selected uniformly at random within the range $[0, 2\pi)$ and similarly repeated every other electrical signal. The duration of all pulses is matched to the clock repetition rate of 1~MHz, i.e. 1~$\mu$s and the final pattern is then repeated every 200~ms.}
\label{f:modulation_signal}
\end{figure}

The experimental set-up depicted in Fig.~3 in the main manuscript, relied on spatial mode sorting of the pseudo thermal states via interferometric means. However, precise and reliable control over the optical path difference of the device was paramount for the desired sorting operation. Therefore, we implemented an active feedback loop mechanism to have direct control over the interferometer and the relative phase difference of its inputs. A reference coherent signal was multiplexed into the input signals via the same EOMs used for the generation of the pseudo thermal states. This mechanism effectively halved the final repetition rate to 500~KHz since every two laser pulses, one was used for calibration and tuning operations. Fig.~\ref{f:modulation_signal} depicts a simplified representation of a one-shot modulation signal used for the phase changing EOM. The first signal applies a voltage that imprints a complete $\pi$ shift onto the coherent state while the second signal applies a random phase uniformly extracted from the set $[0, 2\pi)$ ensuring that the final P-function was not skewed due to limited randomness generation \cite{zanforlin2019, Marangon2016}. The adjustable air-gap placed in one of the optical path of the interferometer was then used to ensure that the phase applied to the reference signal state was kept constant throughout the detectors' integration time thus resulting in high interferometric visibility ($\approx 99\%$).

\section{Optical mask fabrication and characterization}
\begin{figure}
\centering
\includegraphics[width=0.7\linewidth]{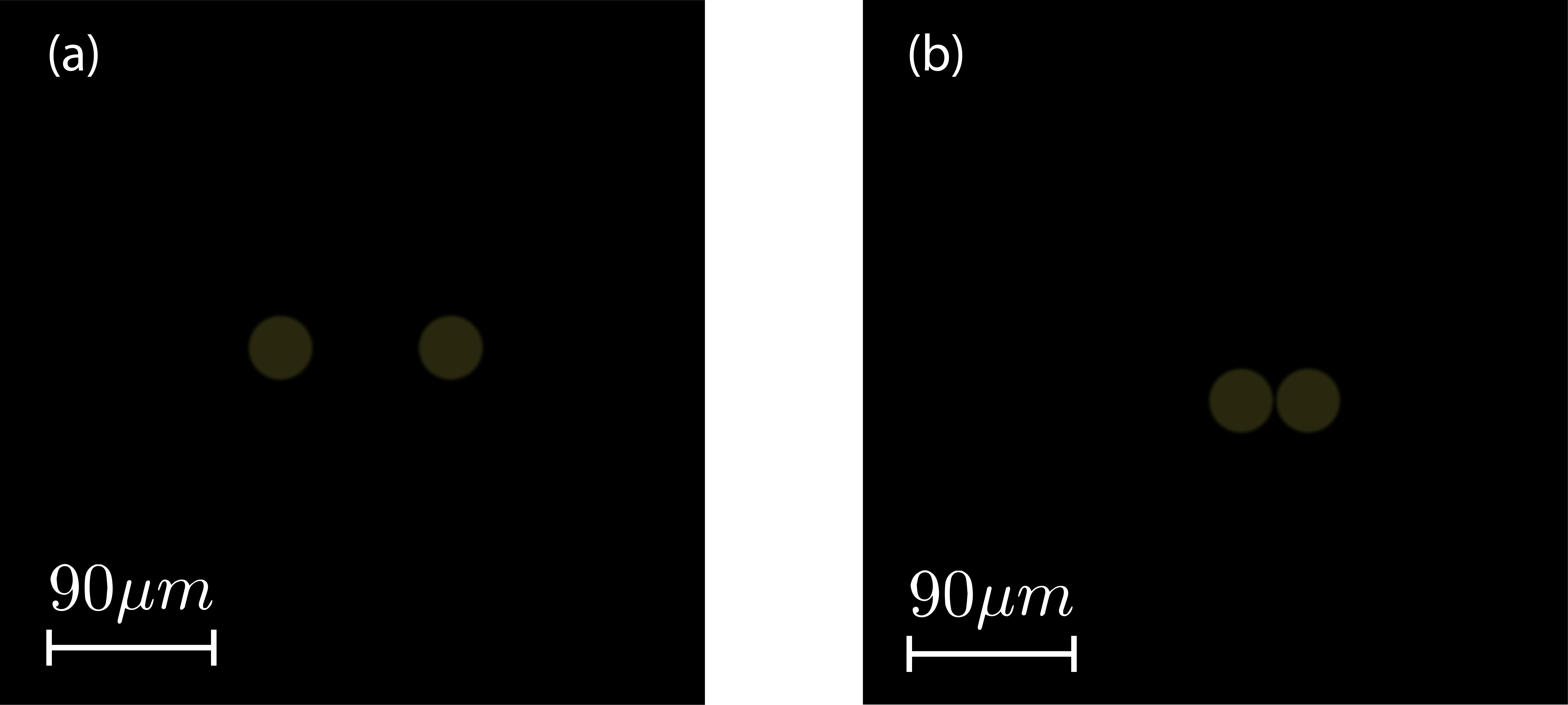}
\caption{Pictures of the final optical masks taken with a Leica DMRM microscope (15x magnification). Image (a) shows a mask with 30~$\mu$m wide pinholes separated by 90~$\mu$m while image (b) shows a mask with pinholes of the same size but a 35~$\mu$m separation. Pictures' contrast has been saturated to better resolve the two pinholes against the background.}
\label{f:masks_microscope_shots}
\end{figure}

The optical masks were formed using patterned etching of thin chromium layers on a fused silica substrate. The 1.5~mm thick fused silica substrates were coated with 90~nm thickness of chromium using an electron-beam vacuum evaporation process forming a layer sufficiently thick to be fully opaque to the near-infrared radiation used in this experiment. The substrate was then coated in 5~nm positive photoresist (\mbox{AZ 1505}), and the pattern (pinholes, reference and alignment markers) inscribed using a Heidelberg DW66+ \mbox{laser-writer}. Once developed, the chromium was removed using a chemical etchant (TechniEtch Cr01). Fig.~\ref{f:masks_microscope_shots} displays pictures of the masks taken with a Leica DMRM microscope (15x magnification).

\begin{figure}
\centering
\includegraphics[width=\linewidth]{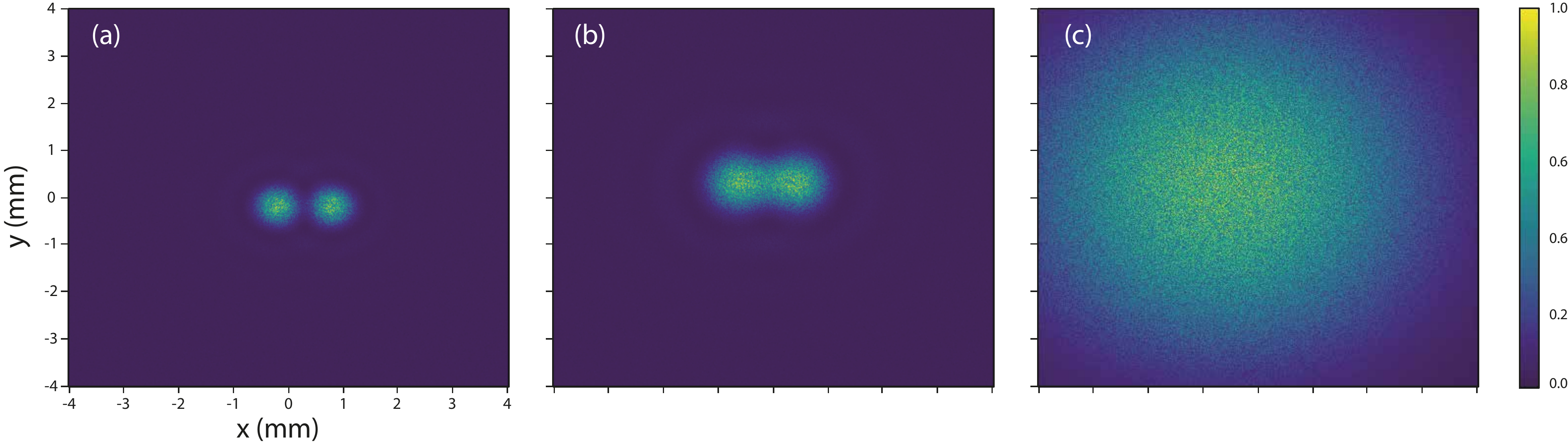}
\caption{Normalised intensity profile pictures of transmitted light through the optical mask. At close imaging distances (a) $z=2$~cm the two pseudo thermal sources are clearly distinct depicting two Airy diffraction patterns with faint secondary rings. As the distance increases, (b) $z=3$~cm the two images merge together until they completely coalesce (c) $z=20$~cm resulting in a heavily diffracted image where it is impossible to distinguish the individual sources. All pictures are relative to the same mask with 30~$\mu$m wide pinholes separated by 1~mm.}
\label{f:int_mask_picts_thm}
\end{figure}

A single-photon sensitive CCD camera (Rolera EM-C\textsuperscript{2} Bio-Imaging Microscopy Camera) was used to extract profile intensity images of the transmitted light by the two circular pinholes at different distances. Fig.~\ref{f:int_mask_picts_thm} shows the normalised intensities at imaging distances $z=2, 3$ and 20~cm for a mask with 30~$\mu$m wide pinholes separated by 1~mm using thermal radiation. At short distances, the intensity profiles depict the classical Airy diffraction pattern expected from circular apertures where faint secondary rings are visible. In these configurations, the two sources can still be resolved, however, as the distance between the masks and the camera increases, diffraction prevails and the profiles merge together removing any knowledge of the initial sources.

\begin{figure}
\centering
\includegraphics[width=0.6\linewidth]{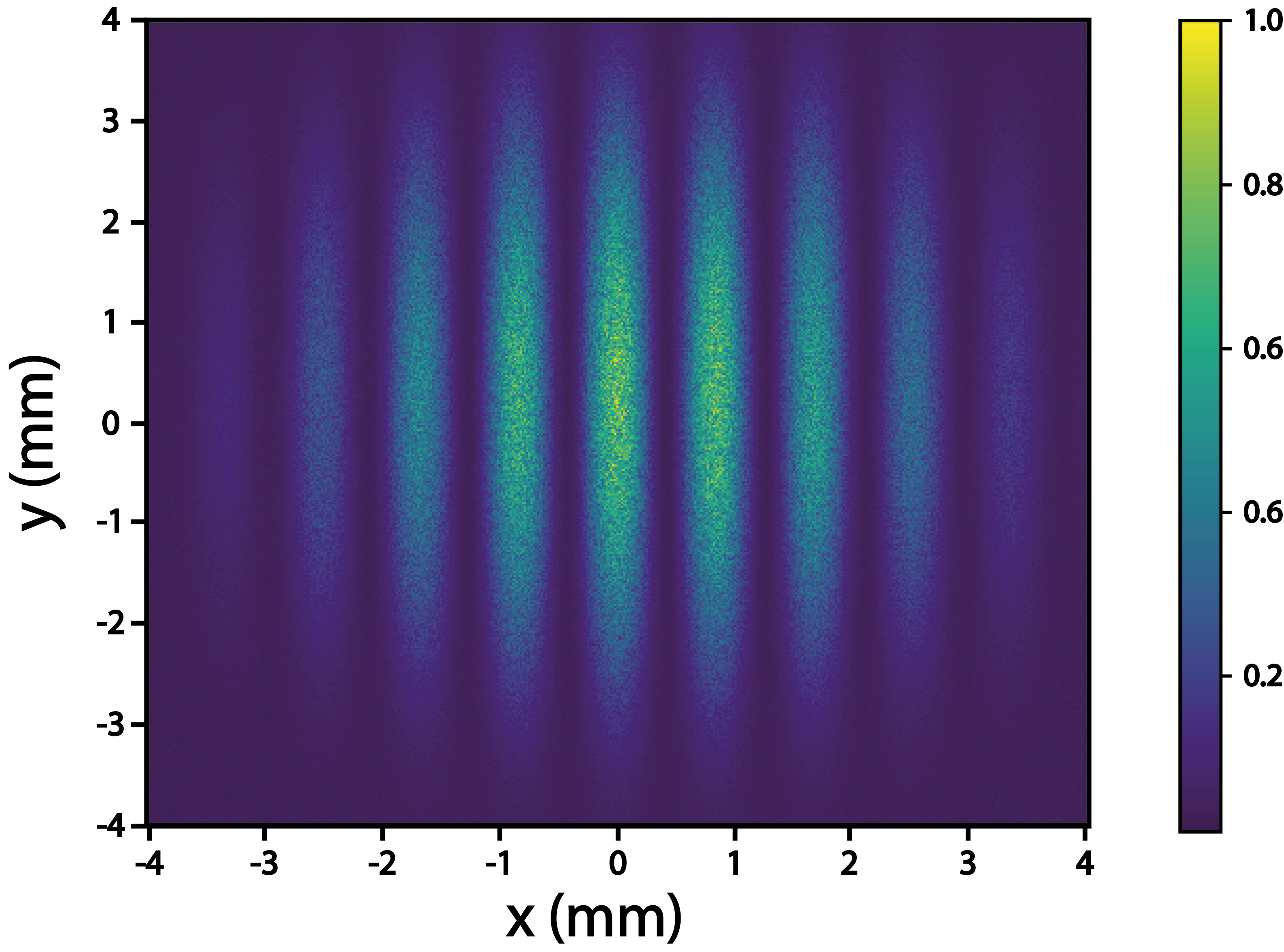}
\caption{Normalised intensity profile picture of transmitted coherent light through the optical mask. Without any phase and intensity modulation provided by the EOMs, the two point-like sources undergo ``classical'' interference as demonstrated by the visible interferometric fringes. The mask used had 30~$\mu$m wide pinholes separated by 150~$\mu$m placed at a distance of 15~cm from the camera. The separation of the fringes is $\approx 0.821$~mm which is consistent with the expected theoretical value for the imaging system used.}
\label{f:int_mask_picts_coh}
\end{figure}

The masks were also tested using coherent radiation to ensure that the thermal generation process successfully removed any spatial correlation that could potentially disrupt the interferometer's mode sorting mechanism. Fig.~\ref{f:int_mask_picts_coh} shows the resulting image for a mask with 30~$\mu$m wide pinholes separated by 150~$\mu$m placed at a distance $z=15$~cm from the camera. Interferometric fringes are clearly recognizable showing a fringe separation of $\approx 0.821$~mm which is in good agreement with the theoretical value expected for this imaging system.

\section{Quantum state discrimination}

In a previous work \cite{PhysRevLett.124.080503}, we have shown that a two-mode interferometer has the same sensitivity in estimating the separation between two sources as SPADE, given comparable numerical apertures. Here we show that the same two-mode interferometer is also optimal in our discrimination problem here. 

\begin{figure}[h!]
\includegraphics[trim = 0cm 0cm 0cm 0cm, clip, width=0.5\linewidth]{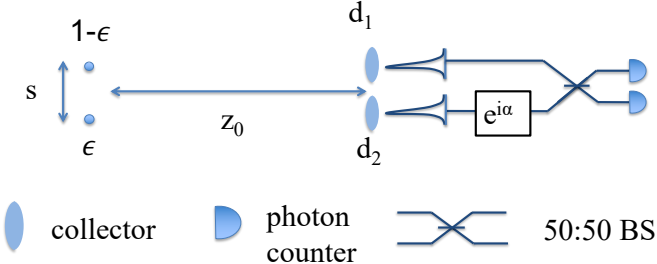}
\caption{\label{supp:f:mzi} Schematic of two sources with a separation of $s$ in the object plane, at a distance $z_0$ from the collectors. Two collectors at $d_1$ and $d_2$ direct light into a two-mode interferometer consisting of a phase shift of $\alpha $  and a 50:50 beam splitter, followed by photon counters.\label{supp:f:mzi}}
\end{figure}

Consider the set-up in Fig.~\ref{supp:f:mzi} where two collectors are places at positions $d_1$ and $d_2$ orthogonal to the optic axis; the collectors are at a distance $z_0$ from the sources. One source (the star) is positioned at $x_0$, and the planet, if present, is positioned at $x_0 + s$.
Assuming we are in the paraxial regime, the optical path difference of the planet between the two collectors is
is

\begin{align} 
\psi_1 
       % &= k\left(z_0 \sqrt{1+\frac{(x_0 + \theta - u_2)^2}{z_0^2} } -
        % z_0 \sqrt{1+\frac{(x_0 + \theta -  u_1)^2}{z_0^2} }
       % \right) \nn
       &\approx k z_0 \left( \frac{(x_0 + s - d_2)^2}{2 z_0^2} - \frac{(x_0 + s -  d_1)^2}{2z_0^2} \right) \nn
       &= k\left(\frac{d_2^2-d_1^2}{2 z_0} + \frac{(d_1 - d_2)(x_0 + s)}{z_0} \right) \nn
\ket{\psi_\text{planet}} &= \frac{1}{\sqrt 2}(\ket{d'_1} + e^{i\psi_1}\ket{d'_2})
\end{align}

The optical path difference of the star to the two collector is
\begin{align}
\psi_2 &= k(\mu_1 - \mu_0) 
 \approx k\left( \frac{d_2^2-d_1^2}{2 z_0} + \frac{(u_1 - u_2) x_0}{z_0}\right) \nn
\ket{\psi_\text{star}} &= \frac{1}{\sqrt 2}(\ket{d'_1} + e^{i\psi_2}\ket{d'_2}) 
\end{align}

The states to discriminate between are
\begin{align}
\rho_a &= \ket{\psi_\text{star}}\bra{\psi_\text{star}} \nn
\rho_b &= (1-\epsilon)\ket{\psi_\text{star}}\bra{\psi_\text{star}} +
 \epsilon \ket{\psi_\text{planet}}\bra{\psi_\text{planet}}
\end{align}

We redefine the angular separation $\theta = s/z_0$.
In the basis of $\rho_a$, the two density matrices are
\begin{align}
\rho_a' &=
 \left(
\begin{array}{cc}
 1 & 0 \\
 0 & 0 \\
\end{array}
\right) \nn
\rho_b'&= \left(
\begin{array}{cc}
 \frac{1}{2} (\epsilon  \cos (\theta k d)-\epsilon +2) & \frac{1}{2} i \epsilon  \sin (\theta k d) \\
 -\frac{1}{2} i \epsilon  \sin (\theta k d) & \epsilon  \sin ^2\left(\frac{1}{2} \theta k d\right) \\
\end{array}
\right),\nn
d &\equiv u_1-u_2
\end{align}

The relative entropy between $\rho_a$ and $\rho_b$ is approximately

\begin{align}
S'(\rho_a||\rho_b) \approx \frac{\theta ^2 k^2 \epsilon  d^2}{4 \log (2)}
\end{align}

Now, we apply the measurement in Fig.~\ref{supp:f:mzi}. We put the collected light at $d_1, d_2$ through a phase shift $e^{i\alpha}$, followed by a 50:50 BS. We
assume the operators transform as
\begin{align}
a'^{\dagger}_{d1} \rightarrow \frac{1}{\sqrt2}(a_{d1}^\dagger + a_{d2} ^\dagger) \nn
a'^{\dagger}_{d2} \rightarrow \frac{e^{i\alpha}}{\sqrt2}(a_{d1}^\dagger - a_{d2} ^\dagger)
\end{align}

For $H_0$, the measurement outcomes are
\begin{align}
p_0(a) = \frac{1}{2}(1+\cos(\psi_1 + \alpha))  \nn
p_0(b) = \frac{1}{2}(1-\cos(\psi_1 + \alpha)) 
\end{align}

For $H_1$, they are 
\begin{align}
p_1(a) = \frac{1}{2}(1-\epsilon)(1+\cos(\psi_1 + \alpha)) + \frac{1}{2}\epsilon(1+\cos(\psi_2 + \alpha)) \nn
p_1(b) = \frac{1}{2}(1-\epsilon)(1-\cos(\psi_1 + \alpha)) + \frac{1}{2}\epsilon(1-\cos(\psi_2 + \alpha)) \nn
\end{align}

Define
\begin{align}
\kappa \equiv \frac{d_2^2-d_1^2}{2 z_0^2}
\end{align}
We set $\alpha$ to
\begin{align}
\alpha = -\kappa +  \frac{\epsilon  (k d (x_0+\theta z_0))}{z_0}+\frac{(1-\epsilon ) (k x_0 d)}{z_0}.
\end{align}
This is analogous to the SPADE method, where we align the apparatus to the weighted center.  That is, this step assumes that we know the centre-of-mass of the two sources. If the planet is absent, then this is equal to the position of the star.

The classical relative entropy of this measurement is
\begin{align}
S(p_0||p_1)= &\cos ^2\left(\frac{1}{2} d \theta k \epsilon \right)  \left(\log \left(\cos ^2\left(\frac{1}{2} d \theta k \epsilon \right)\right)-\log \left(\frac{1}{2} (\epsilon  \cos (d \theta k (\epsilon -1))-(\epsilon -1) \cos (d \theta k \epsilon )+1)\right)\right)\nn
&+\sin ^2\left(\frac{1}{2} d \theta k \epsilon \right) \left(\log \left(\sin ^2\left(\frac{1}{2} d \theta k \epsilon \right)\right)-\log \left(\frac{1}{2} (-\epsilon  \cos (d \theta k (\epsilon -1))+(\epsilon -1) \cos (d \theta k \epsilon )+1)\right)\right) \nn
\approx & \frac{\theta^2 k^2 \epsilon d^2}{4 \log (2)},
\end{align}
\noindent which is optimal in the limit that $\theta,\epsilon \ll 1$ (see Supp. Material).

\section{Angular separation estimation}

\subsection{The reference laser}

In the experiment, the laser is input into a multimode fibre, goes through free space (sprays out like a spherical point source), and then comes through the two slits.
Since there is no randomness introduced in the process, we model the output at the two slits as a plane wave of coherent states.
The reference laser provides a calibration for the applied phase $\alpha$.

\begin{figure}[h]\center
\includegraphics[trim = 0cm 0cm 0cm 0cm, clip, width=0.6\linewidth]{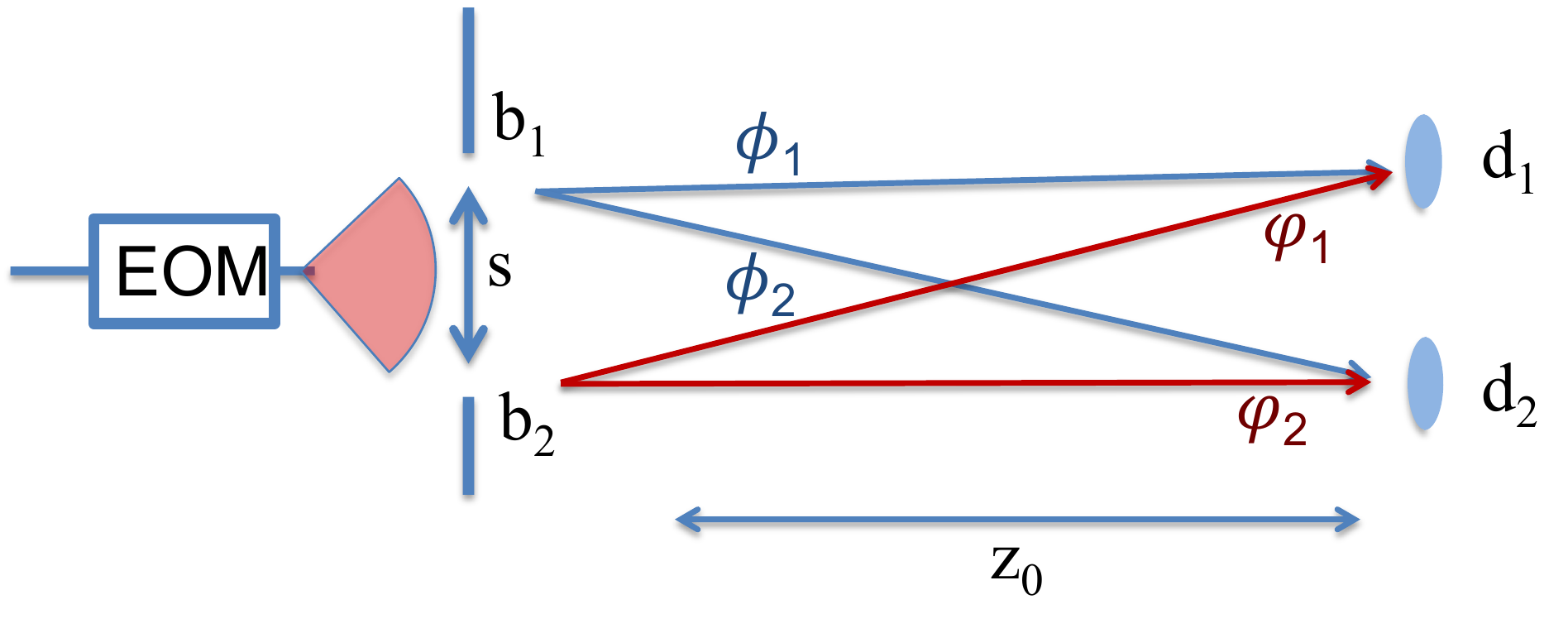}
  \caption{Schematic figure for part of the experimental setup.\label{supp:f:simple_scheme} }
\end{figure}

The electric field at $d_1$ is given by the combination of the two fields of the two souces
\begin{align}
\hat E_{d1} =  \sqrt{\eta}(\hat E_{b_1}e^{-i \phi_1} + \hat E_{b_2} e^{-i \psi_1})\nn
\hat E_{d2} =  \sqrt{\eta}(\hat E_{b_1}e^{-i \phi_2} + \hat E_{b_2} e^{-i \psi_2})\nn
\end{align}

The photon numbers at $d_1$ and $d_2$ are:
\begin{align}
|E_{d1}|^2 &= \eta (\hat E^*_{b_1}e^{+i \phi_1} + \hat E^*_{b_2} e^{i \psi_1})(\hat E_{b_1}e^{-i \phi_1} + \hat E_{b_2} e^{-i \psi_1}) \nn
% &= \eta( |E_{b1}|^2 + E^*_{b1} E_{b2}e^{i(\phi_1 - \psi_1)}  + E^*_{b2} E_{b1}e^{i(\psi_1 - \phi_1)})  + |E_{b2}|^2) \nn
&= \eta( |E_{b1}|^2 + |E_{b2}|^2 + 2|E_{b1}|^2 \cos( -\phi + \kappa) ) \nn
|E_{d2}|^2 &= \eta(\hat E^*_{b_1}e^{i \phi_2} + \hat E^*_{b_2} e^{i \psi_2})
                  (\hat E_{b_1}e^{-i \phi_2} + \hat E_{b_2} e^{-i \psi_2}) \nn
           % &=\eta( |E_{b_1}|^2 + |E_{b2}|^2 + E_{b_1}^* E_{b_2} e^{i(\phi_2 - \psi_2)} + E^*_{b_2} E_{b1} e^{i(\psi_2 -\phi_2)} ) \nn
           &=\eta( |E_{b_1}|^2 + |E_{b2}|^2 + 2 |E_{b1}|^2 \cos(\phi + \kappa) )\nn
           % &= \eta( )
\end{align}

The optical path differences are 
\begin{align}
\phi_2 -\phi_1 &= \phi, \quad \psi_2 - \psi_1 = \psi = -\phi \nn
\phi_1 - \psi_2 &\approx 0, \quad \phi_2 -\psi_1 \approx  0
\end{align}

Since the two slits are of the same size (equal intensity), we model the coherent states at $b_1$ and $b_2$ as 
$$
\rho_{b1}= \ket{\beta}\bra{\beta}, \qquad  \rho_{b2} =\ket{\beta e^{i \kappa}}\bra{\beta e^{-i \kappa}}.
$$
 
We need to calculate the expectation value of this observable to obtain the correlations:
\begin{align}\label{eq:correlations}
\hat O = & \text{Re}[E^*_{d1} E_{d2} e^{i \alpha} ] = \frac{1}{2}(E^*_{d1} E_{d2} e^{i \alpha} + cc. )
\nn
      = & \frac{\eta}{2}\left[ (E^*_{b1}e^{i \phi_1} + \hat E^*_{b2} e^{i \psi_1})
                              (E_{b_1}e^{-i \phi_2} + \hat E_{b_2} e^{-i \psi_2})e^{i\alpha}  + c.c. \right]\nn
= & \frac{\eta}{2}\bigg\{ e^{i\alpha}\left[|E_{b1}|^2 e^{i(\phi_1 - \phi_2)} + E^*_{b1} E_{b2} e^{i(\phi_1 - \psi_2)} +
E^*_{b2} E_{b1}e^{i(\psi_1 - \phi_2)} + |E_{b2}|^2 e^{i(\psi_1 -\psi_2)} \right]  + \nn
&e^{-i\alpha}\left[|E_{b1}|^2 e^{-i(\phi_1 - \phi_2)} + E_{b1} E^*_{b2} e^{-i(\phi_1 - \psi_2)} +
E_{b2} E^*_{b1}e^{-i(\psi_1 - \phi_2)} + |E_{b2}|^2 e^{-i(\psi_1 -\psi_2)} \right]  \bigg\}\nn
    % = &\eta\left[ |E_{b1}|^2 \cos(-\phi + \alpha)  + |E_{b2}|^2\cos(-\psi + \alpha)
       % + (E_{b1}^* E_{b2} + E_{b1} E^*_{b2}) \cos(\alpha) 
       % + (E_{b1}^* E_{b2} + E_{b1} E^*_{b2}) \cos(\alpha)    \right] 
& = \eta (|E_{b_1}|^2 \cos(\phi_1 - \phi_2 + \alpha)      
   + (E^*_{b_1 } E_{b_1} e^{i(\phi_1 - \psi_2 + \alpha + \kappa)} 
     +E_{b_1} E^*_{b_1}e^{-i (\phi_1 -\psi_2 +\alpha + \kappa)} ) + \nn
     & (E^*_{b_1} E_{b_1}e^{i(\psi_1 - \phi_2 +\alpha -\kappa)} +
        E_{b_1} E^*_{b_1}e^{-i(\psi_1 - \phi_2 +\alpha -\kappa) }) +
 |E_{b_2}|^2 \cos(\psi_1 - \psi_2 + \alpha)                
\end{align}
The expectation value of the operator in Eq.~\eqref{eq:correlations}
is 
\begin{align}
\braket{O } =\eta\left[\beta^2 \cos(-\phi + \alpha) + \beta^2 \cos(-\phi+\alpha)  + 4 \beta^2\cos(\kappa+\alpha)) \right]
\end{align}

After the collectors ate $d_1$ and $d_2$, we have a phase-shifter and 50:50 beam splitter. This transformation gives
\begin{align}
\hat E_{d1} &\rightarrow (\hat A + \hat B )/\sqrt 2 \nn
\hat E_{d2} &\rightarrow  e^{-i\alpha}(\hat A - \hat B )/\sqrt 2
\end{align}
\noindent 

We can now calculate the statistics of the coherent state at the two detectors, $A$ and $B$.
Inversing the above gives

\begin{align}
A &= \frac{1}{\sqrt2}(E_{d1} + e^{i\alpha }E_{d_2}) \nn
B &= \frac{1}{\sqrt2}(E_{d1} - e^{i\alpha }E_{d_2}) \nn
A ^* A &=\frac{1}{2}(E^*_{d1} + e^{-i\alpha }E^*_{d2})(E_{d1} + e^{i\alpha }E_{d2}) \nn
           &= \frac{1}{2}\left(|E_{d1}|^2 + |E_{d2}|^2 + 2\text{Re}[ E^*_{d1} E_{d2} e^{i \alpha} ] \right) \nn
B ^* B &= \frac{1}{2}(E^*_{d1} - e^{-i\alpha }E^*_{d_2}) (E_{d1} - e^{i\alpha }E_{d_2})\nn
           &= \frac{1}{2}(|E_{d1}|^2 + |E_{d2}|^2 - 2\text{Re}[ E^*_{d1} E_{d2} e^{i \alpha} ]   ) \nn
N_\text{total} &= A ^* A + B ^* B
\end{align}

If the interferometer is imperfect, where some of the signal is replaced by noise, we model this as

\begin{align}
A^*_{ \text{noisy }} A_{ \text{noisy }} &= \nu A ^* A + (1-\nu) N_\text{total}/2 \nn
B^*_{ \text{noisy }} B_{ \text{noisy } }&= \nu B ^* B + (1-\nu) N_\text{total}/2 
\end{align}

If we were to compute the difference in photon counts between the two detectors, we find that
\begin{align}\label{eq:ratio}
\mathcal{R} &= \frac{A^*_{ \text{noisy }} A_{\text{noisy }} - B^*_{ \text{noisy }} B_{\text{noisy }} }{N_\text{total}} \nn
&=\frac{(1-\nu) \cos (\alpha ) (\cos (\kappa )+\cos (\phi ))}{\cos (\kappa ) \cos (\phi )+1}\nn
&= \nu\cos(\alpha) \quad \text{when $\kappa = 0$}
\end{align}

For $\kappa = 0$, which is our case here, since the two slits are equidistant from the output of the multimode fibre, the above expression reduces to $\cos\alpha$. That is, the reference laser behaves almost like a single point source.
 The parameter $\kappa$ being non-zero will only reduce the visibility of the measurement, and the effect is almost neglible. 
 
We measure the parameter $\nu\cos(\alpha)$ directly from experimental data, which is then used to update the probability distribution in the maximum likelihood method.

\subsection{Maximum likelihood}
In our analysis, we use a maximum likelihood method to obtain an estimator.
Using Eq.~11 in the main text, we can estimate $|\phi|$ by the estimator $\phi_\text{}$, then obtain $\theta= s/z_0$ from
\begin{align}
\hat \theta_\text{est} = 2\hat \phi_\text{est}/(k d)
\end{align} 

We can use maximum likelihood method to obtain $|\phi|$ via Baye's theorem.
Therefore, we would like to obtain the probability distribution for $|\phi|$, $\mathcal{P}(\phi)$ given the detection events, $\alpha$ and $\mathcal{R}$.
For two events $A$ and $B$, Baye's theorem states that
\begin{align}
P(A|B) = \frac{P(B|A)P(A)}{P(B)}.
\end{align}

Initially the probability distribution for $\mathcal{P}(\phi)$ is uniform in $[0,2\pi]$, therefore $\mathcal{P}_0(\phi) = 1/(2\pi)$. After one detection event $\mu = a,b$ and adjustable phase $\alpha$, we have
\begin{align}
\mathcal{P}(\phi|\mu,\alpha,\nu) \propto  \mathcal{P}(\mu|\phi,\alpha,\nu) \mathcal{P}_0(\phi|\alpha,\nu),
\end{align}
Here there is a normalisation factor that can be fixed `easily'.
We know $\mathcal{P}(\mu|\phi)$, these are
\begin{align}\label{papb}
\mathcal{P}( a|\phi,\alpha,\nu) = \frac{1}{2}\left( 1+ \nu \cos(\alpha)\cos\left[ \phi \right] \right)\nn
\mathcal{P}( b|\phi,\alpha,\nu) = \frac{1}{2}\left( 1- \nu \cos(\alpha)\cos\left[ \phi \right] \right)
\end{align}

Given a detection event $\mu=a,b$ where the adjusted phase was $\alpha$, the probability for $\phi$ can be updated  via
\begin{align} \label{eq:one_event}
\mathcal{P}(\phi|\mu,\alpha,\nu) \propto \mathcal{P}(\mu|\phi,\alpha,\nu) P_0
\end{align}
Therefore, after the detection event, Eq.~\eqref{eq:one_event} is updated using 
Eq.~\eqref{papb}, depending on whether $a$ or $b$ occurred. After $m$ detection events, we have the vector of deteciton events, e.g. $\vec \mu_m = (a,b,b,a,...)$, given a vector of adjustable phases $\vec \alpha = (\alpha_1,\alpha_2,...,\alpha_m)$. In the experiment, $\alpha$ is constant for each data point.

We have
\begin{align}\label{eq:bayes}
\mathcal{P}(\phi|\vec \mu_m, \vec \alpha,\nu) \propto 
P(\mu|\phi, \alpha,\nu)\mathcal{P}(\vec \mu_{m-1}| \phi, \vec \alpha,\nu)
\end{align}

Now, since all the functions we deal with here are sinusoid, we can conveniently express them as a Fourier series.
After $m$ clicks, the probabilities can be expressed as a Fourier series
\begin{align}\label{eq:fourier}
P(\phi|\vec \mu_m, \vec \alpha) = \frac{1}{2\pi} \sum_{k = -m} ^m a_k e^{i k \phi}
\end{align}
\noindent where $m$ here corresponds to the higher order of the Fourier coefficient.
The coefficient of the term $e^{i k \phi}$ is denoted $a_k$, which depends on $\vec \mu_m $ and $\alpha_m$. Normalising $\alpha_0$ to $1/2\pi$ will keep the entire distribution normalised.

\begin{figure}[h!]\center
\includegraphics[trim = 0cm 0cm 0cm 0cm, clip, width=0.45\linewidth]{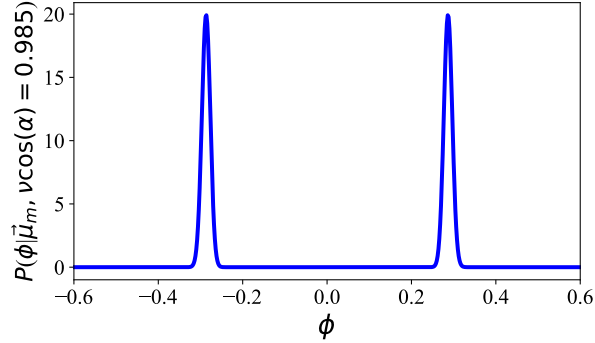} 
  \caption{PDF for $\phi$, after 10000 detection events which 206 were from detector $b$  }\label{supp:f:pdf}
\end{figure}

 We equivalently write the update events in this Fourier form.
 For example, if detector $b$ fires, then
\begin{align} \label{eq:update}
\mathcal{P}(\mu=b|\phi,\alpha,\nu) = \frac{1}{2}[1 - \nu \cos (\alpha) \cos(\phi) ].
\end{align}
We write Eq.~\eqref{eq:update} as
\begin{align}
\mathcal{P}(\mu=b|\phi,\alpha,\nu) = \frac{1}{2} - \frac{1}{4} \nu \cos \alpha  e^{i \phi}  - \frac{1}{4} \nu \cos \alpha  e^{- i \phi}
\end{align}
therefore the update coefficients are $a_0 =\pi, a_1 = a_{-1} = \frac{\pi}{2} \nu \cos (\alpha)$.
This example is particularly relevant, because the factor $\nu \cos (\alpha)$ is equal to $\mathcal{R}$ in Eq.~\eqref{eq:ratio}, and is directly measured in the experiment using the calibration laser.

Before the first detection, Eq.~\eqref{eq:fourier} only contains one term, $a_0=1$. After each detection event given by the probabilities in Eq.~\eqref{papb}, the number of Fourier coefficients grow by 2; $a_k$ are updated using Eq.~\eqref{eq:bayes}, which once again uses Eq.~\eqref{papb}.

After the coefficients are obtained, $P(\phi)$ can be evaluated for each value of $\phi$, and we extract the maximum. There are two peaks for $\phi$ of equal intensity, since $\cos(\phi) = \cos(-\phi).$
We show a simulated example in Fig.~\ref{supp:f:pdf}, we use the parameter $\nu \cos\alpha = 0.985$; after 10000 detection events where $206$ were were output at detector $b$.

\end{document}